# mRNA-miRNA Reciprocal Regulation Enabled Bistable Switch Directs Cell Fate Decision

Xiao-Jun Tian[1], Hang Zhang[2], Jingyu Zhang[1], and Jianhua Xing[1,3]

1 Department of Computational and Systems Biology, School of Medicine, University of Pittsburgh, Pittsburgh, PA, 15260, USA
2 Genetics, Bioinformatics, and Computational Biology Program, Virginia Polytechnic Institute and State University, Blacksburg, VA, 24060, USA
3 Computational Science Research Center, Beijing, 100193, China

Correspondence:

X.-J. Tian and J. Xing, Department of Computational & Systems Biology, School of Medicine, University of Pittsburgh, 3501 Fifth Avenue, Pittsburgh, PA 15213

Phone: 412-648-0174 (X.-J. Tian)

Phone: 412-383-5743 (J. Xing)

Fax: 412-648-3163 (X.-J. Tian and J. Xing)

Emails: xjtian@pitt.edu, xing1@pitt.edu

**Abstract**

miRNAs serve as crucial post-transcriptional regulators in various essential cell fate decision. However, the contribution of the mRNA-miRNA mutual regulation to bistability is not fully understood. Here, we built a set of mathematical models of mRNA-miRNA interactions and systematically analyzed the sensitivity of response curves under various conditions. First, we found that mRNA-miRNA reciprocal regulation could manifest ultrasensitivity to subserve the generation of bistability when equipped with a positive feedback loop. Second, the region of bistability is expanded by a stronger competing mRNA (ceRNA). Interesting, bistability can be emerged without feedback loop if multiple miRNA binding sites exist on a target mRNA. Thus, we demonstrated the importance of simple mRNA-miRNA reciprocal regulation in cell fate decision.

**INTRODUCTION**

MicroRNAs (miRNAs) are small non-coding RNA molecules containing about 22 nucleotides that exist ubiquitously in many living organisms for post-transcriptional regulation of gene expression. Growing studies reveal that miRNAs are essential in various essential cell fate decision processes such as pluripotency and reprogramming [1], epithelial-to-mesenchymal transition (EMT) [2], cancer stem cells [3] and metastasis [4]. Cells make a choice between two alternative fates: either with high/low level of mRNA/miRNA or with the opposite expression pattern. Dysregulation of miRNAs correlates with pathological conditions such as cancer development, cardiovascular and metabolic diseases. During the last decade, studies have accumulated on the basic molecular mechanisms of miRNA biogenesis, function and degradation [5]. With recent quantitative measurements on miRNA dynamics using techniques such as quantitative fluorescence microscopy [6], it becomes a timely and urgent need to perform systematic mathematical analysis on the mRNA-miRNA mutual regulation and its effect on gene regulatory network dynamics.

Through base-pairing interactions, miRNA inhibits its target mRNA through two modes, translational repression and mRNA degradation [7]. The degree of sequence complementarity between miRNA and mRNA determines the mode of mRNA silencing [8]. Extensive complementarity, which are often formed in plants, induces cleavage and degradation of the target mRNA [9, 10]. Partial complementarity, which occurs between the vast majority of miRNAs and their target mRNA in metazoans, results in translational repression or degradation [7]. In addition, under some circumstance, miRNA can stimulate translation of mRNA through an Argonaute/FMR1-mediated mechanism [11]. Interestingly, miRNA can establish a threshold of target mRNA [6].

Reversely, mRNA targets reciprocally control the stability and function of miRNAs [11-13]. Target interaction can stabilize an miRNA by preventing its release from Argonaute (Ago) and subsequent destabilization [14]. The miRNAs in a mRNA-miRNA complex may be either degraded together with the mRNA with extensive complementarity [15-17], or be recycled [18]. Furthermore, each miRNA may target tens to hundreds mRNAs [19, 20], enabling cross-talk between competing endogenous RNAs (ceRNAs) targeted by the same miRNA [21]. Consequently, the reciprocal regulation between miRNAs and their targets adds a significant level of complexity to the mRNA-miRNA relationships.

Given the prevalent involvement of mRNA-miRNA interaction in the cell fate decision processes, extensive efforts have been made to determine the thermodynamic standard free energy of binding between mRNAs and miRNAs $\Delta G^0$, and several computational tools are available for *in silico* prediction [19]. Figure 1A shows the distribution of standard free energy of binding between miR-34a and mRNAs of 354 human gene calculated using PicTar [22]. Notably different mRNAs may have the same $\Delta G^0$. Figure 1B gives four such examples. These mRNAs, with only one miR-34a binding site, form complexes with miR-34a with drastically different configurations and number of complementary base pairs. It is questionable that these mRNAs, even under the same condition such as concentrations of involved molecular species, undergo the same miR-34a mediated regulation kinetics. On the other hand, miR-34a forms different positive feedback with its targets, such as miR-34/snail1 [23], miR-34a/SIRT1/p53 [24], miR-34a/IL-6R/STAT3 [25] (Fig. 1C). Moreover, these mRNA-miRNA feedback loops play important roles on cell fate decisions, such as miR-34/snail1 in the partial epithelial-to-mesenchymal transition [26, 27]. However, the contribution of the ultrasensitivity from mRNA-miRNA reciprocal regulation to the bistability is controversial [27-32]. Thus, it is urgent to explore the critical roles of the mRNA-miRNA mutual interaction on the cell fate decision in a general model.

In this work we use mathematical and computational analysis to demonstrate the critical role of the mRNA-miRNA mutual interaction on cell fate decision. We found that the reciprocal regulation between mRNA and miRNA is either ultrasensitive or subsensitive, and either inhibitive or protective. Ultrasensitivity from the mRNA-miRNA reciprocal regulation contributes to bistability generation when it is equipped with a positive feedback loop. Furthermore, the region of bistability is expanded when a stronger competitor (ceRNA) is involved since the degree of response sensitivity is amplified. Alternatively, bistability can be generated from mRNA-miRNA reciprocal interaction when there are more than one miRNA binding sites on the target mRNA.

## MODEL AND METHODS

### Model of mRNA-miRNA reciprocal regulation

Figure 2A summarizes all possible scenarios of mRNA-miRNA reciprocal regulation. miRNAs either suppress an mRNA through translational repression or accelerated degradation, or activate an mRNA through stimulated translation. Reversely, an mRNA either suppresses an miRNA through accelerated degradation, or activates an miRNA through sequestering it from degradation. In order to analyze the contribution of the ultrasensitivity from the mRNA-miRNA reciprocal regulation on the generation of bistability, an inhibition arm in which the protein product of mRNA inhibits the synthesis of miRNA can be added to enclose a double-negative feedback loop (Fig. 2B). Figure 2C shows the corresponding kinetic schemes for an mRNA with $N$ miRNA binding sites. There are $C_N^i = N!/(i!(N-i)!)$ different complexes that consist of one mRNA and $i (\leq N)$ copies of miRNA.

Noticing that the binding/unbinding events between miRNAs and mRNAs are typically much faster than other processes, such as transcription, translation and degradation, we assume that the binding/unbinding processes can be approximated to be in quasi-equilibrium. With no detailed information on cooperativity of mRNA-miRNA binding, we assume that each miRNA binds to the mRNA independently, and the binding free energy for each binding site is the same ($\Delta G^0$). Then each form of the mRNA-miRNA$_i$ complex ($R_i$) has the same level, denoted as $[R_i]$, and the total level of the mRNA-miRNA$_i$ complex is $C_N^i [R_i]$. Furthermore, the levels of free miRNA and mRNA ($[miR]$, $[mR]$), the total levels of miRNA and mRNA ($[miR]_t$, $[mR]_t$) and the mRNA-miRNA$_i$ complex ($[R_i]$) are constrained as below,

$$[mR] = [mR]_t - \sum_{i=1}^{N} C_N^i [R_i], \qquad (1)$$

$$[miR] = [miR]_t - \sum_{i=1}^{N} i\, C_N^i [R_i]. \qquad (2)$$

Under the quasi-equilibrium approximation, the following relationship exists between $[R_i]$ and $[R_{i-1}]$:

$$K_i [miR][R_{i-1}] = [R_i], \qquad i = 1 \ldots N, \qquad (3)$$

where $[miR]$ is the cellular level of the free miRNA under study, $[R_0]$ is defined as the mRNA concentration $[mR]$, and $K = \exp(-\Delta G^0)$ is the binding constant. Here $\Delta G^0$ is in the unit of $k_B T$, the product of Boltzmann's constant and temperature. The degradation rate constant of $[R_i]$ is $d_{Ri}$, and the translation rate of the $[R_i]$ is $k_{si}$. Upon degradation of the complex mRNA-miRNA$_i$, miRNA molecules can be recycled with a ratio $\lambda_i$ ($0 \le \lambda_i \le 1$). Thus, the equations of total level of miRNA, mRNA and protein ($[P]$) are:

$$\frac{d[miR]_t}{dt} = \frac{k_{miR}}{(1 + F * ([P]/J)^n)} - d_{miR}[miR]$$
$$- \sum_{i=1}^{N}(1-\lambda_i) i C_N^i d_{Ri}[R_i], \qquad (4)$$

$$\frac{d[mR]_t}{dt} = k_{mR} - d_{mR}[mR] - \sum_{i=1}^{N} C_N^i\, d_{Ri}[R_i], \qquad (5)$$

$$\frac{d[P]}{dt} = k_{s0}[mR] + \sum_{i=1}^{N} k_{si}\, C_N^i [R_i] - d_p[P]. \qquad (6)$$

For convenience of following discussions, we reform the above equations as,

$$\tau_{miR} \frac{d[miR]_t}{dt} = \frac{\alpha_{miR}}{(1 + F * ([P]/J)^n)} - [miR]$$

$$- \sum_{i=1}^{N} (1 - \lambda_i) i C_N^i \beta_{Ri}[R_i], \tag{7}$$

$$\tau_{mR} \frac{d[mR]_t}{dt} = \alpha_{mR} - [mR] - \sum_{i=1}^{N} C_N^i \beta_{Ri}[R_i], \tag{8}$$

$$\tau_P \frac{d[P]}{dt} = \alpha_{P0}[mR] + \sum_{i=1}^{N} \alpha_{Pi} C_N^i [R_i] - [P]. \tag{9}$$

where, $\tau_{miR} = 1/d_{miR}$, $\tau_{mR} = 1/d_{mR}$, $\tau_P = 1/d_P$, $\alpha_{miR} = k_{miR}/d_{miR}$, $\alpha_{mR} = k_{mR}/d_{mR}$, $\beta_{Ri} = d_{Ri}/d_{mR}$, $\alpha_{P0} = k_{s0}/d_p$, $\alpha_{Pi} = k_{si}/d_p$. The value $\tau_{miR}$, $\tau_{miR}$ and $\tau_{miR}$ indicate the time scale of miRNA, mRNA and Protein respectively, the value of $\alpha_{miR}$, $\alpha_{mR}$, and $\alpha_{P0}$ determine the level of miRNA, mRNA and Protein without mutual regulation respectively, while the value of $K_i$, $\beta_{Ri}$, $\lambda_i$ and $\alpha_{Pi}$ indicates the strength of the inhibition between each other. The parameters values are used as follows except specified. $\alpha_{miR} = 1$, $\alpha_{mR} = 1$, $\alpha_{P0} = 10$, $\alpha_{Pi} = 1$, $\beta_{Ri} = 5$, $K_i = 10 \sim 100$, $\lambda_i = 0.5$. Parameter $\alpha_{miR}$ and $\alpha_{mR}$ will used for dose response curve analysis while the affection of other parameters on the sensitivity analysis will also systematically analyzed. It is noted that the time scale does not change the steady state. So the exact value of $\tau_{miR}$, $\tau_{mR}$ and $\tau_P$ do not affect the conclusion, we used $\tau_{miR} = 1$, $\tau_{mR} = 1$, $\tau_P = 1$. Throughout this work all the variables and parameters are in arbitrary units.

**Bistability Analysis**

To describe the inhibition of miRNA synthesis by the protein, an inhibitory Hill function ($\frac{J^n}{J^n + [P]^n}$) is multiplied to $\alpha_{miR}$ as shown in Eqn. (7). The analysis with this positive feedback can be achieved by setting $F = 1$ instead of $F = 0$ in the case of without positive feedback. If the Hill coefficient $n \leq 1$, this inhibition arm is not ultrasensitive. Therefore, in this model system the required nonlinearity can come either from protein mediated inhibition on miRNA synthesis, or directly from the mRNA-miRNA mutual regulation. To show the contribution of these two sources of ultrasensitivity to bistability, the minimum of Hill coefficient ($n_{min}$) to generate bistability will be analyzed. If $n_{min} < 1$, then the bistable system required ultrasensitivity can result from mRNA-miRNA mutual interaction. The method of finding $n_{min}$ is shown in Fig.S1. First, one-parameter bifurcation diagram is analyzed to check the bistability and find the Saddle-Node bifurcation points. Then based on these two Saddle-Node bifurcation points, two-parameter bifurcation diagram is analyzed to find the cusp bifurcation point, where $n_{min}$ is located.

**Sensitivity Analysis**

The steady-state response curve is often used to describe how the output of the system (O) depends on the input (I). To quantify the sensitivity of the system, the instantaneous sensitivity is defined as the ratio of the fractional changes in response output ($\Delta O/O$) and stimulus input ($\Delta I/I$),

$$s(I) = \lim_{\Delta I \to 0} \frac{\Delta O/O}{\Delta I/I} = \frac{dO/O}{dI/I} = \frac{dlog(O)}{dlog(I)}.$$

Instantaneous sensitivity is also known as 'logarithmic gain' in biochemical systems theory [33] or as 'local sensitivity coefficient' in the local parameter sensitivity analysis [34]. The response is ultrasensitive if $|s| > 1$, subsensitive if $0 < |s| < 1$, desensitive if $s = 0$ and linear if $|s| = 1$. The sign of $s$ indicates whether $I$ inhibits or activates $O$. In general, the instantaneous sensitivity $s$ is not constant but depends on the input ($I$). We denote the dependence of instantaneous sensitivity $s$ on $I$ as the instantaneous sensitivity curve, and the extremum of this curve as the maximum (in the sense of the absolute value) sensitivity ($s_m$) of the specific I-O curve. Under this definition, the maximum sensitivity of a Hill function ($\frac{I^n}{I^n + K^n}$) is exactly the Hill coefficient $n$. Thus, $s_m$ can be used as the gauge of the degree of sensitivity. The steady-state response curves and bifurcation diagrams are performed with PyDSTool [35].

**RESULTS**

The major aim of this paper is to explore the contribution of ultrasensitivity from mRNA-miRNA mutual regulation to the generation of bistability. We first did sensitivity analysis for mRNA-miRNA mutual regulation with single binding site, then analyzed the contribution of the ultrasensitivity to bistability with a feedback loop. We also explored the effect of the ceRNA on the bistability generation. Last, we extend our analysis to multiple binding sites.

**The ultrasensitivity from mRNA-miRNA mutual regulation subserves the generation of bistability when equipped with a positive feedback loop.**

Given that miRNA and mRNA molecules are able to reciprocally regulate each other, we analyzed the contribution of the sensitivity from the mutual regulation on the generation of bistability and cell fate decision. Let's start with the case that an mRNA has only one binding site for the miRNA.

First, the sensitivity of regulation of mRNA/protein by miRNA is analyzed. As shown in Figure 3A, under mode of absence of feedback regulation, the protein concentration decreases with the miRNA synthesis rate constant $\alpha_{miR}$, reflecting inhibition of miRNA on mRNA/protein. Most of the response curves are sigmoidal shaped with respect to $\alpha_{miR}$ in logarithmic scale, exhibiting a progression from weak inhibition that accelerates and approaches total inhibition with increase of $\alpha_{miR}$. The sharpness of the sigmoidal shape decreases and eventually disappears when the recycle ratio $\lambda$ increases. Shape changes of the response curves are also reflected by the corresponding instantaneous sensitivity curves (Fig. 3B), which are bell-shaped with an extremum in the middle, where the mRNA level is most sensitive to miRNA level change. The maximum sensitivity $s_m$, which is negative for inhibition, increases with the recycle ratio $\lambda$ (Fig. 3C). That is, the larger the recycle ratio, and so the more efficient of miRNA, the less sensitive the inhibition of mRNA by miRNA is. Therefore, ultrasensitivity is generated by sacrificing the efficiency of inhibiting the mRNA. Notably, when the recycle ratio is near 1, $|s_m|$ is less than 1. That is, regulation of mRNA by miRNA shows subsensitivity instead of ultrasensitivity when the miRNA is almost completely recycled.

A positive feedback loop with an ultrasensitivity arm can produce bistability [36], which plays essential roles in cell fate decision. We hypothesize that ultrasensitivity from the mRNA-miRNA reciprocal regulation contributes to the generation of bistability. Thus, the response curves under the presence of feedback regulation is analyzed. Figure 3D clearly shows the existence of bistable regions of the protein level while varying $\alpha_{miR}$ even with Hill coefficient $n = 1$. The existence of bistability is also demonstrated in Fig. S2, the nullclines of $[mR]_t$ and $[miR]_t$ have three intersection points, two of which are stable steady states and the other is an unstable steady state. The two-parameter bifurcation diagrams in Fig. 3E show how the parameter region of bistability changes over $n$ and $\alpha_{miR}$. As expected, with other parameters fixed the value of $n$ needs to exceed a critical value to generate bistability. This critical value has cusp-shaped dependence on $\alpha_{miR}$, which relates to the bell-shaped sensitivity curves in Fig. 3B. Notably this critical value of $n$ can be less than 1 for certain values of $\alpha_{miR}$. That is, both the mRNA-miRNA mutual regulation and the protein regulation on miRNA synthesis contribute to the generation of bistability, as long as the composite nonlinearity exceeds a threshold value.

It is noted that the bistable region decreases with the recycle ratio. That is, the miRNA recycle ratio $\lambda$ also affects the critical value of Hill coefficient $n_{min}$. Indeed Fig. 3E shows that the space of $\lambda$ and Hill coefficient $n$ is divided into monostable and bistable regions. The smaller the value of $\lambda$, the smaller the Hill coefficient $n$ that is required for generating bistability. This is consistent with the results that the maximum sensitivity ($s_m$) of the reciprocal regulation between miRNA and mRNA increases with the decrease of recycle ratio (Fig. 3C). That is, when miRNA is almost completely recycled, a larger nonlinearity from the other source is required to generate bistability. In general, when the miRNA is not fully recycled, bistability can come from synergistic effect of the mRNA-miRNA reciprocal regulation and protein-miRNA transcriptional regulation. This conclusion is also confirmed by exploring the dependence of $s_m$ and $n_{min}$ on changing of other parameters (Fig. S3). For example our previous study shown that a double negative feedback loop between Snail1 and miR-34 functions as a bistable switch to control the transition of epithelial to partial EMT state [26, 27]. In this system snail1 mRNA has only one binding site for miR-34 [23], and the nonlinearity can come from both Snail1 inhibition of miR-34 transcription (with $n = 2$) and miR-34/snail1 mutual regulation in human [27]. While in rat/mouse, the nonlinearity comes from miR-34/snail1 mutual regulation since only one Snail1 binding site exists on the miR-34 promoter (with $n = 1$) [23]. That is, the ultrasensitivity from miRNA-mRNA mutual regulation and that from Hill coefficient can compensate each other to generate bistability.

The threshold dynamics is more transparent in Fig. 3G. When plotted against the mRNA synthesis rate constant, $\alpha_{mR}$, the protein level remains low until $\alpha_{mR}$ exceeds a threshold value so that mRNA molecules escape from miRNA-mediated repression by titrating the miRNA in the system, and the threshold increases with $\lambda$. The threshold behavior is consistent with the quantitative measurements of Mukherji et al [6] (yellow circles in Fig. 3G).

Furthermore, the sensitivity of miRNA regulated mRNA dynamics depends on the mechanistic details, which can be translational repression stimulation, degradation or of translation. Figure 3H shows the maximum sensitivity $s_m$ in the parameter space spanned by the degradation rate constant $\beta_{R1}$ and translation rate constant $\alpha_{P1}$ of the mRNA-miRNA complex. In the blue region located in right-bottom corner, $s_m < -1$, thus regulation of mRNA by miRNA shows different degree of ultrasensitivity. Thus, in this region, small $n_{min}$ is enough to generate bistability (Fig 3I). In the green region $-1 < s_m < 0$, thus only subsensitivity can be obtained and $n_{min}$ need to be larger than 2. In this region, $n_{min}$ may need to be larger than 2 to generate bistability. In the orange region, $s_m > 0$, the regulation of mRNA by miRNA shows subsensitive stimulation, which is largely from the stimulation of translation of mRNA by forming complex with miRNA. It is noted that, in subsensitive stimulation region, bistability cannot be generated. Taken together, the regulation of mRNA by miRNA can be ultrasensitive or subsensitive inhibition, and subsensitive activation and the ultrasensitivity from the regulation of mRNA by miRNA can absolutely subserve the generation of bistability and cell fate decision.

**Competing endogenous RNAs amplify ultrasensitivity of mRNA-miRNA mutual regulation and thus extend the bistable region**

In the above analyses, the miRNA has only one target mRNA. Actually, miRNA typically targets multiple mRNAs [19, 37]. For example, miR-200 targets both zeb2 and pten [38]. To examine the effect of ceRNAs on the regulation of mRNA/Protein by miRNA, we expanded the model to include a ceRNA.

As shown in Fig. 4A, without ceRNA, $[mR]_t$ first decreases slowly with increasing $\alpha_{miR}$, then the decrease accelerates after $\alpha_{miR}$ exceeds a threshold value so more miRNA molecules are available to effectively inhibit mRNA. The response curve of $[P]$ changes at the presence of ceRNA, and the change depends on the value of $K_c$, the ceRNA-miRNA binding constant. When $K_c \ll K_1$, ceRNA only manifests its effect on the level of $[P]$ at large $\alpha_{miR}$ when its concentration is sufficiently high to compete with the target mRNA. On the other hand, when $K_c \gg K_1$, at small values of $\alpha_{miR}$ the target mRNA under interest is largely not affected by the miRNA inhibition since the ceRNA binds most of the miRNA molecules, then its amount drops suddenly after $\alpha_{miR}$ is sufficiently large so the miRNA molecules titrate out the ceRNA molecules. Consequently, the $\alpha_{miR}$-$[P]$ response curve first decreases then increases its sharpness with $K_c$, which is revealed by the $s_m(\alpha_{miR}\text{-}P) - K_c$ curve in Fig. 4B. That is, when the ceRNA-miRNA binding constant is small, ceRNA desensitizes the regulation of the mRNA by miRNA. However, after the binding constant exceeds a threshold, ceRNA sensitizes the regulation.

The above modulation of sensitivity, however, depends on the recycle ratios $\lambda$ and $\lambda_c$. As shown in Fig. 4E, depending on $K_c$, the pattern of $s_m(\alpha_{miR}\text{-}P)$ in the space of $\lambda$ and $\lambda_c$ changes qualitatively. With $K_c < K_1$, for a fixed $\lambda_c$, $|s_m|$ decreases with increased value of $\lambda$, consistent with that in Fig. 3. For a fixed $\lambda$ though, $|s_m|$ increases with $\lambda_c$. Especially when $\lambda_c \to 1$, the response becomes significantly more sensitive to miRNA inhibition. With $K_c = K_1$, $s_m(\alpha_{miR}\text{-}P)$ shows a linear anti-correlated dependence on $\lambda$ and $\lambda_c$. This linear dependence turns out to be specific only for using the same parameter sets for the mRNA and the ceRNA except the recycle ratios, otherwise the results would be qualitatively similar to the case of $K_c < K_1$ or $K_c > K_1$.

With $K_c > K_1$, the qualitative feature of the response is roughly the opposite of that with $K_c < K_1$, while the absolute value of $s_m$ is much bigger. For a fixed $\lambda_c$, $|s_m|$ initially decreases monotonically with an increasing value of $\lambda$, then has an extremum at an intermediate value of $\lambda$ when $\lambda_c \to 1$. For a fixed $\lambda$, $|s_m|$ instead decreases with $\lambda_c$. Despite its difficulty to perform simple mathematical analyses, these results demonstrate that presence of a ceRNA further complicates the dynamic behavior of the mRNA-miRNA regulation.

Intuitively, a ceRNA provides certain protection of the target mRNA from miRNA inhibition through competitive binding of the latter. Indeed Fig. 4D shows that $[P]$ increases with the ceRNA synthesis rate constant $\alpha_{mRc}$, and shows positive sensitivity. Figure 4E shows that the sensitivity of this curve increases with $K_c$. Regulation of mRNA by ceRNA is subsensitive when ceRNA binding is weak (compared to $K_1$), and ultrasensitive when ceRNA binding is strong. The response curve sensitivity also depends on the recycle ratio $\lambda$. As shown in Fig. S4A, $s_m(k_{mRc}$-$[P])$ increases with $\lambda$ and $K_c$. Interestingly, the value of $s_m$ does not change with $\lambda_c$ except at the complete recycle point (Fig. S4B). Therefore, mRNAs can regulate each other indirectly by acting as a ceRNA of the other. This miRNA-mediated crosstalk tunes the sensitivity of mRNA-miRNA regulation. Several studies exist on the molecular determinants of effective ceRNA crosstalks [39-44], such as miRNA/ceRNA ratio, numbers of total and shared miRNA response elements, and target binding affinity. However, the functionality of this increased sensitivity from ceRNA is not studied before. Thus, the minimum of Hill coefficient $n_{min}$ for bistability is also analyzed by considering a ceRNA (Fig. 4F-G). $n_{min}$ also first increases then decreases with $K_c$. That is, a ceRNA with high binding affinity can also contribute to the generation of bistability in cell fate decision system. This is consistent with the non-monotonic dependence of $s_m$ on $K_c$ (Fig. 4B). Taken together, competing endogenous RNAs is able to amplify ultrasensitivity of mRNA-miRNA mutual regulation and thus extend the bistable region.

**Multiple miRNA binding sites on target mRNA lead to ultrasensitivity and bistability**

In the above analyses we modeled mRNAs with one miRNA binding site. Some mRNAs have multiple miRNA binding sites. For example, miR-200 can target to zeb1/2 mRNA on 5-6 highly conserved binding sites [45]. Thus, we systematically examined how the existence of two miRNA binding sites affect the mRNA-miRNA mutual regulation and the generation of bistability.

Since there are two binding sites, there are two groups of possible mRNA-miRNA complexes, either with one miRNA bound ($R_1$) or two miRNAs bound ($R_2$) on mRNA. We denoted the recycle ratios of two mRNA-miRNA complexes as $\lambda_1$ and $\lambda_2$, respectively. Figure 5A shows how the total mRNA level depends on the miRNA synthesis rate $\alpha_{miR}$ under different combinations of $\lambda_1$ and $\lambda_2$. Similar to the case of one binding site, the $\alpha_{miR}$-$[P]$ curve also shows ultrasensitivity and subsensitivity (cyan and magenta lines respectively in Fig. 5A). More interestingly, different from the response curve with one binding site, bistability occurs under appropriate combination of $\lambda_1$ and $\lambda_2$ (black line in Fig. 5A). To the best of our knowledge, there is no previous report on the bistability that results from the mutual regulation between miRNA and mRNA without any feedback loop. Future experiments can test this novel prediction by fine tuning the two recycle ratios.

To further explore how to generate specific response by tuning the recycle ratios, we locate bistable and monostable domains in the space of $\lambda_1$ and $\lambda_2$. As shown in Fig. 5B, bistability is generated in the corner where $\lambda_1$ is much smaller than $\lambda_2$ (blank region), while in other region, different level of sensitivity is generated. It is noted the bistable-monostable boundary depends on other parameters. As shown in Fig.S5, the bistable regions can be expanded by increasing the mRNA-miRNA binding constant. The plausible underlying mechanism of bistability generated with multiple binding sites is that formation of mRNA-miRNA$_2$ protects miRNA under large $\lambda_2$, while an increased level of miRNA promotes formation of mRNA-miRNA then mRNA-miRNA$_2$ with ultrasensitivity under smaller recycle ratio $\lambda_1$ (Fig. 3). This is equivalent to an inherent positive feedback loop with ultrasensitivity, meeting the requirement of bistability [36]. The reason why bistability cannot be generated from one-binding site model is that the protective effect and ultrasensitivity cannot coexist simultaneously with one recycle ratio.

We further considered the involvement of the positive feedback loop and analyzed the contribution of mRNA-miRNA mutual regulation with multiple binding sites to bistability. We analyzed the dependence of $n_{min}$ on the recycle ratios $\lambda_1$ or $\lambda_2$ respectively with the other as a constant. First, with $\lambda_2$ fixed at 0.8 and decrease of $\lambda_1$, the bistable region increases as the cusp point gradually decreases as shown in the two-parameter bifurcation diagrams (Fig. 5C). Notably as $\lambda_1$ is small enough, the curve intersects with Hill coefficient boundary $n = 0$. This is because that bistability can be generated directly from mRNA-miRNA mutual regulation with multiple binding sites without a positive feedback loop as we demonstrated in Fig .5A-B. We use $n_{min} = 0$ to represent this intrinsic bistability. Over all, $n_{min}$ increases with $\lambda_1$ (as shown in Fig. 5D). This is consistent with the dependence of $n_{min}$ on recycle ratio in the case with one binding site (Fig. 3F). However, the dependence of $n_{min}$ on $\lambda_2$ is completely opposite. As shown in Fig. 5E-F, with constant $\lambda_1 = 0.8$, $n_{min}$ decreases with $\lambda_2$. In addition, the effect of ceRNA on miRNA-mRNA mutual regulation is analyzed in the cases with two binding sites on mRNA or ceRNA respectively in Fig. S6 and Fig. S7 respectively. In both cases, $n_{min}$ first increases then decreases with ceRNA-miRNA binding constant, showing a similar trend as Fig.4F-G. That is, a strong ceRNA also could expand the bistability region under the case with multiple binding sites.

In summary, the reciprocal regulation between mRNA and miRNA is more versatile with two binding sites than that with one binding site. Bistable switch has a more possibility to be existed in the mRNA-miRNA system with multiple binding sites. It is interesting to generalize similar analysis on the mRNA-miRNA systems with more than two binding sites, such as miR-200/zeb1/2 system in which 5-6 binding sites are highly conserved [45].

**DISCUSSION**

It is generally suggested that a regulatory miRNA serve as rheostats to fine-tune the expression of its targets to accommodate cell responses [46, 47]. Here, we systematically analyzed mutual regulation between miRNA and mRNA with a class of basic mathematical models. To our surprise, we found that this reciprocal regulation gives rise to rich dynamic features, which plays important roles in regulating cellular processes such as cell phenotype change and maintenance.

**mRNA-miRNA regulation provides a new mechanism for ultrasensitivity and bistability**

The dynamics of a biological system is typically nonlinear. The nonlinearity results from a large variety of sources [48, 49], such as cooperativity [50], homo-multimerization [51], zero-order ultrasensitivity [52], multi-site phosphorylation [53, 54], substrate competition [55], lateral interaction [27] and molecular titration [28]. Bistable and multi-stable switches are often involved in cell fates decision in biological systems [26, 56, 57]. Reports showed that miRNA regulation of mRNAs contributes to the robustness of biological system when considering feedback or feedforward loops [58-60]. Indeed, our analyses showed that miRNA/mRNA reciprocal regulation is also a source of nonlinearity. It shows ultrasensitive and subsensitive inhibition, and subsensitive protection. This nonlinearity can contribute to the generation of bistable switch and cell fate decision in a biological system.

Positive feedback and ultrasensitivity are two prerequisites for bistability. Here we showed that bistability is generated when ultrasensitivity from the mRNA-miRNA reciprocal regulation is equipped with a positive feedback loop. Furthermore, bistability also can be generated from mRNA-miRNA reciprocal regulation when more than one miRNA binding sites exist in the absence of any imposed feedback regulation. This result is analogous to the bistability from multi-site phosphorylation [61, 62]. However, the underlying mechanisms of bistability are different for multisite mRNA-miRNA reciprocal regulation and multi-site phosphorylation. The former results from coexistence of ultrasensitivity and miRNA protection, while the later arises from substrate saturation and competitive inhibition [61].

**mRNA-miRNA regulation provides mechanism for pathway crosstalks**

Since there are typically multiple targets for a single miRNA, mRNAs can cross-regulate each other by competing for the shared miRNAs [38, 39, 43]. Thus, miRNAs link individual signaling pathways and form an intertwined ceRNA network (ceRNETs) [63]. Taking the ceRNETs into account makes the overall regulatory network more complex and leads us to rethink our view of the design principles of biological networks. Coupling of our model with other bio-networks such as protein-protein interaction network, metabolic network, gene regulatory network will give more interesting dynamic phenomena. In addition, design of miRNA sponges should take into consideration of the potential effect of ceRNAs on a gene regulation network with miRNA involved. Here, we found that bistable region is expanded because larger degree of ultrasensitivity can be achieved when a stronger ceRNA is involved. Thus, one can take advantage of this feature to design more effective miRNA sponges.

**Thermodynamic and kinetic parameters control dynamic features of mRNA-miRNA regulation**

The versatile dynamics of mRNA-miRNA regulation results from interplays among a number of thermodynamic and kinetic parameters. For example, in this work we discussed how varying the miRNA recycle ratio alone can qualitatively change the regulation dynamics. It is already shown that miRNAs can shield from exonucleolytic degradation through interacting with its target [14, 64]. The underlying molecular mechanism is not well understood. It is suggested that target mRNA keeps miRNA bound with the Argonaute protein, protecting it from degradation [14].

Furthermore, it is implied that the extent of protection is positively correlated with the number of available target sites [11, 64]. Here, we showed how this molecular-level detail manifests itself in the context of the network-level mRNA-miRNA regulation dynamics.

**Quantitative measurements can test model predictions**

It is surprising that a simple mRNA-miRNA motif, with only one or two miRNA binding sites and possibly presence of just one ceRNA, can generate such diverse dynamics by varying a few controlling parameters. Further studies may likely reveal even richer dynamics when the module is placed in the context of the global network dynamics and cell regulation, where multiple RNAs with different numbers of miRNA binding sites compete for common miRNAs. Given the varying copy numbers of both miRNAs and mRNAs [37], the consequence of stochasticity, which is not discussed here, is another important topic to be explored. An open question is whether cells have explored and utilized all these theoretical possibilities, or functional requirements have converged the parameters to specific regions of the multi-dimensional parameter space.

To address the above question, it is necessary to have quantitative characterization of both thermodynamic and kinetic parameters of mRNA-miRNA regulation. The task is challenging but has been carried out at certain extent. For example, it has been experimentally shown that miRNA is multiple-turnover, enabling several rounds of target recognition and cleavage per miRNA [18, 65, 66]. Mathematically, the number of the round of target recognition and cleavage per miRNA, $N_{round}$, is related to the recycle ratio $\lambda$, satisfying $N_{round} = 1/(1-\lambda)$. Therefore one can estimate $\lambda$ from the measured $N_{round}$. For example, each human let-7-containing complex directs on average 10 rounds of mRNA cleavage [65], while each human miR-233 molecule regulates at least 2 target mRNA molecules [18], then the corresponding values of $\lambda$ are 0.9, and 0.5, respectively. Haley and Zamore showed that a let-7 siRNA-directed ribonucleoprotein complex catalyzes more than 50 target RNA cleavages [66], $i.e.$, $\lambda > 0.98$. The extent of the complementary between miRNA and mRNA may also provide a clue on the value of $\lambda$. That is, the free energy of mRNA-miRNA binding is another factor to control the recycle ratio. Then one can design a specific miRNA with certain recycle ratio according the degree of complementary to achieve desired sensitivity. Combined computational structural modeling and experimental efforts may further accelerate the process.

The mRNA-miRNA regulation dynamics needs to be explored at the level of network dynamics. Mukherji et al. showed experimentally the ultrasensitivity of miRNA-mediated regulation of mRNA [6], which results from molecular titration [28]. Similar quantitative measurements, together with synthetic designs that can modify various thermodynamic and kinetic parameters, can test the predictions made in this work, such as the tunability of ultrasensitivity by controlling the recycle ratio. It is noted that polymorphism in miRNAs and their target sites may give more variation for the miRNA-mRNA binding and thus the recycle ratio. Integration of the information from PolymiRTS database [67] with our model analysis can give further insight on functional variation of miRNA in physiological and pathological phenotype.

In summary, mRNA-miRNA mutual regulation can generate versatile dynamics. Detailed understanding of the mechanistic details and functional roles of the regulation requires consorted efforts from both quantitative measurements and computational analyses at both molecular and system levels. As mRNA-miRNA regulation prevalently exists in various biological systems and some of them function as a 'hub' to coordinately regulates others [68], the versatile effects of mRNA-miRNA reciprocal regulation is evolved to adapt to specific biological context and is robust to intrinsic and extrinsic noise. Thus, understanding the functionality of mRNA-miRNA in the biological system is critical for learning its underlying design principle.

## Acknowledgments

We thank Dr. Bing Liu for helpful discussions. The research was supported by the U.S. National Science Foundation (Grant DMS-1462049).

## Author contributions

X.-J.T. and J.X. conceived and designed the study, X.-J.T. performed the simulation, X.-J.T., H.Z., J.Z. and J.X. analyzed the data, X.-J.T. and J.X. wrote the paper, H.Z. and J.Z. helped with constructive discussion. All authors read and approved the manuscript.

**Figure Caption**

**Fig.1** Target prediction of miR-34a using PicTar [22]. (A) Distribution of standard free energy of binding between miR-34a and targeting mRNAs. (B) Examples of predicted mRNA-miRNA complex configurations for different target mRNAs with the same $\Delta G^0$ but different number of complementary base pairs. (C) Examples of miR-34a mediated positive feedback loops.

**Fig.2** Mathematical model setup. (A) Reported possible mRNA-miRNA reciprocal regulations. (B) Positive feedback motif, in which miRNA and mRNA regulate each other through base-pairing interactions while the protein product of mRNA inhibits the transcription of miRNA by binding to its promoter. (C) Kinetic schemes for mRNA-miRNA interaction with multiple miRNA binding sites in the sequence of mRNA. A positive feedback loop can be formed if Protein inhibits the transcription of miRNA.

**Fig.3** Sensitivity and bistability analyses of the mRNA-miRNA reciprocal regulation, with one miRNA binding site on mRNA. (A) Dependence of the protein level on the miRNA synthesis rate constant $\alpha_{miR}$ under different values of the recycle ratio $\lambda$ without feedback loop. (B) Dependence of the instantaneous sensitivity on $\alpha_{miR}$ under different values of $\lambda$. (C) Dependence of the maximum sensitivity $s_m$ of the $\alpha_{miR}$-$[P]$ response curve on the value of $\lambda$. (D) Dependence of the protein level on the miRNA synthesis rate constant $\alpha_{miR}$ under different values of the recycle ratio $\lambda$ with feedback loop. Dotted line denotes unstable states. (E) The bistable region in the space of $\alpha_{miR}$ and Hill coefficient $n$ under different value of $\lambda$. (E) Dependence of the min Hill coefficient $n_{min}$ for bistability under different values of $\lambda$. (G) Dependence of the protein level on the expression rate constant of mRNA $\alpha_{mR}$ under different recycle ratio $\lambda$, also plotted are experimental data (yellow circles) from ref [6]. (H) The maximum sensitivity $s_m$ of the $\alpha_{miR}$-$[P]$ response curve in the parameter space spanned by the mRNA-miRNA complex translation rate constant $\alpha_{P1}$ and the degradation rate constant $\beta_{R1}$. (I) The minimal Hill coefficient $n_{min}$ for bistability in the parameter space of $\alpha_{P1}$ and $\beta_{R1}$. In the blank region, either $n_{min}$ is larger than 2 or no bistability is found.

**Fig.4** Competitions between different mRNAs for a common type of miRNA affects the mRNA sensitivity to miRNA inhibition and bistability. Each mRNA is modeled with one miRNA binding site. (A) Response curves of the Protein level ($[P]$) on the miRNA synthesis rate constant $\alpha_{miR}$, under different values of the ceRNA-miRNA binding constant $K_c$. The dashed line is the case without ceRNA. (B) The maximum sensitivity $s_m$ of the $\alpha_{miR}$-$[P]$ response curve depends on the ceRNA-miRNA binding constant $K_c$. (C) Dependence of the maximum sensitivity $s_m$ of the $\alpha_{miR}$-$[P]$ response curve on recycle ratios of the miRNA upon degradation of mRNA-miRNA complex ($\lambda$) and the ceRNA-miRNA complex ($\lambda_c$) under different ceRNA-miRNA binding constant. (D) Response curves of the Protein level ($[P]$) on the ceRNA synthesis rate constant, $\alpha_{mRc}$. (E) The maximum sensitivity $s_m$ of the $\alpha_{mRc}$-$[P]$ curve depends on the ceRNA-miRNA binding constant. (F) Two-parameter bifurcation diagrams for $\alpha_{miR}$ v.s. $n$

respectively with different ceRNA-miRNA binding constant $K_c$. (G) The dependence of $n_{min}$ on ceRNA-miRNA binding constant $K_c$. $\lambda_c = 0.5$, $\alpha_{mRc} = 1$, $\beta_{mRc1} = 10$.

**Fig.5** Ultrasensitivity and bistability from multiple miRNA binding sites on one mRNA molecule. (A) Response curves of $[P]$ on the miRNA synthesis rate constant $\alpha_{miR}$ under different combinations of the miRNA recycle ratios upon degradation of complex $mNRA\text{-}miRNA$ ($\lambda_1$) and $mRNA\text{-}miRNA_2$ ($\lambda_2$). Dotted line denotes unstable states. (B) The maximum sensitivity $s_m(\alpha_{miR}\text{-}[P])$ in the space of $\lambda_1$ and $\lambda_2$. The space is divided into bistable (blank region) and monostable (with different sensitivity indicated by heatmap). (C) Two-parameter bifurcation diagrams for $\alpha_{miR}$ v.s. $n$ with different levels of $\lambda_1$ under constant $\lambda_2 = 0.8$. (D) The dependence of $n_{min}$ on $\lambda_1$. (E) Two-parameter bifurcation diagrams for $\alpha_{miR}$ v.s. $n$ with different levels of $\lambda_2$ under constant $\lambda_1 = 0.8$. (F) The dependence of $n_{min}$ on $\lambda_2$. $d_{mR2} = 10$ and $K_1 = K_2 = 100$.

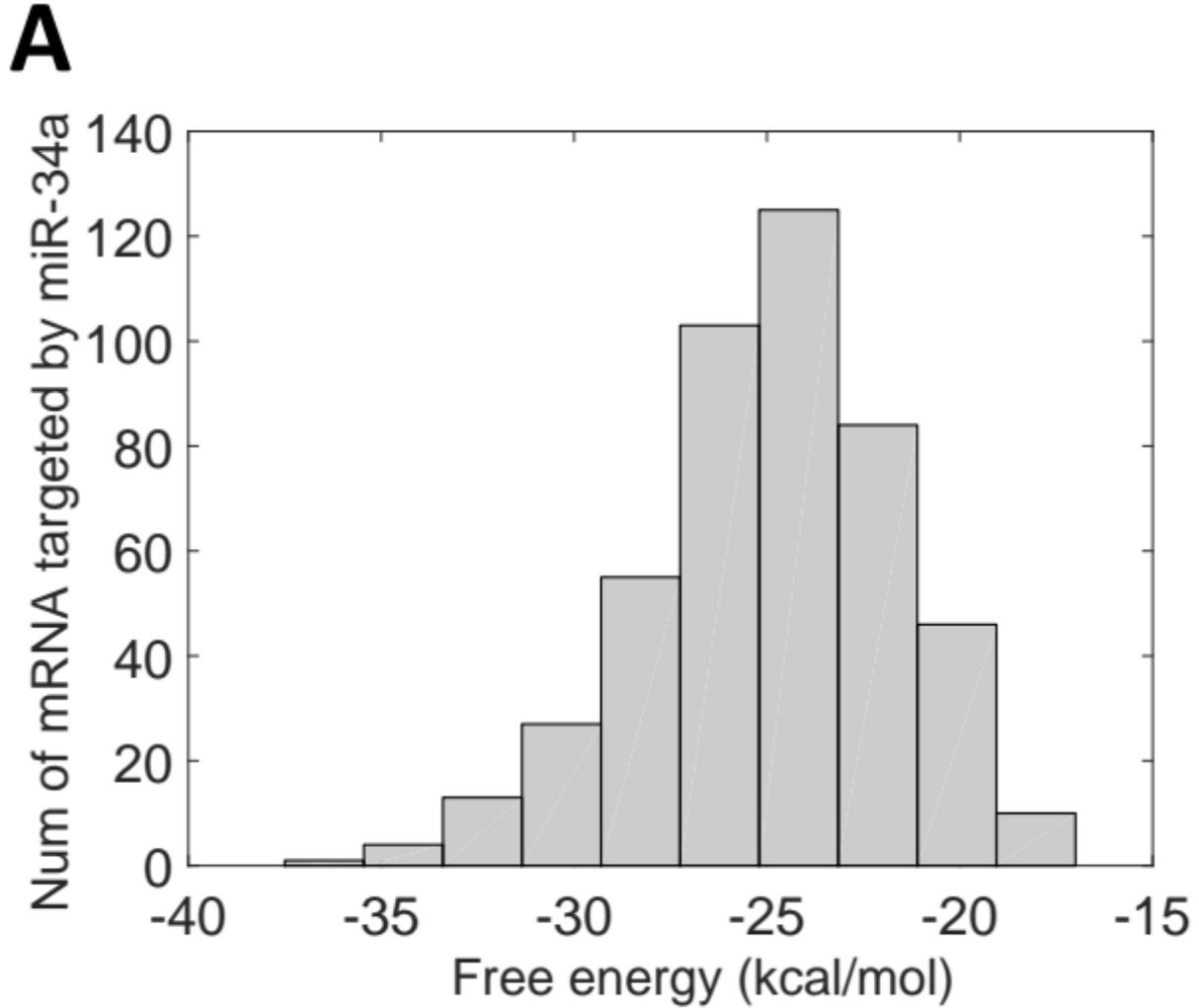 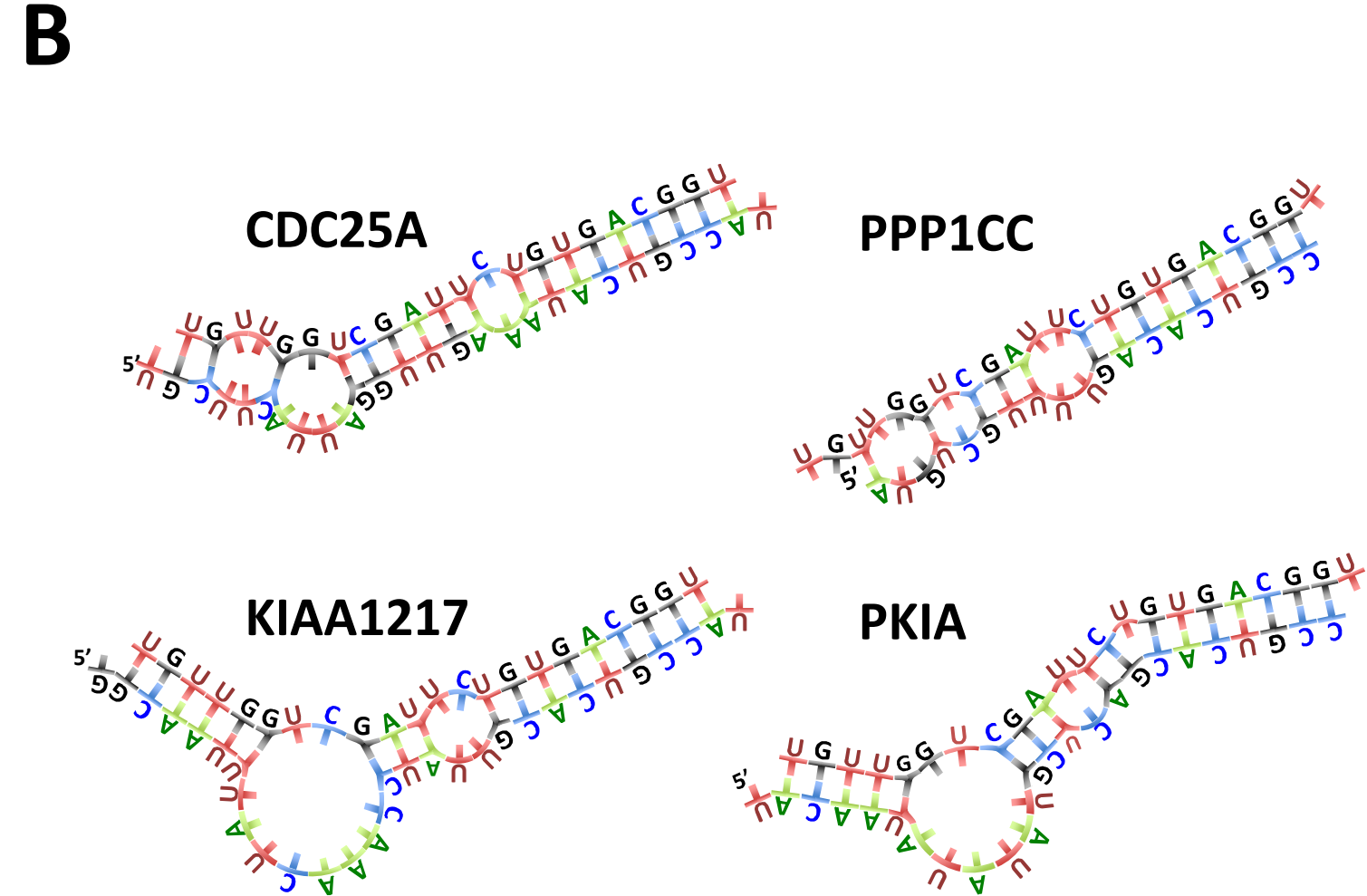 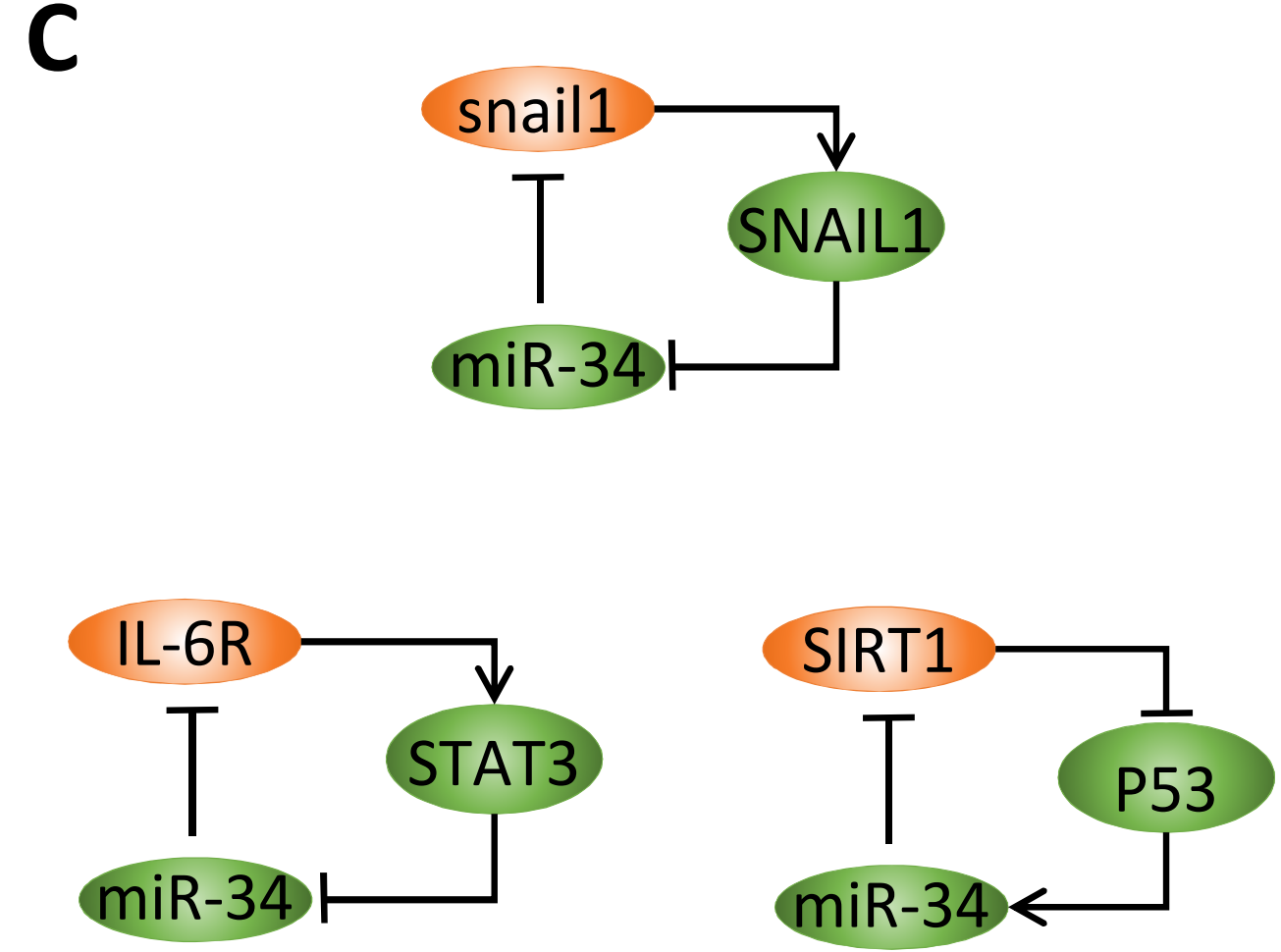

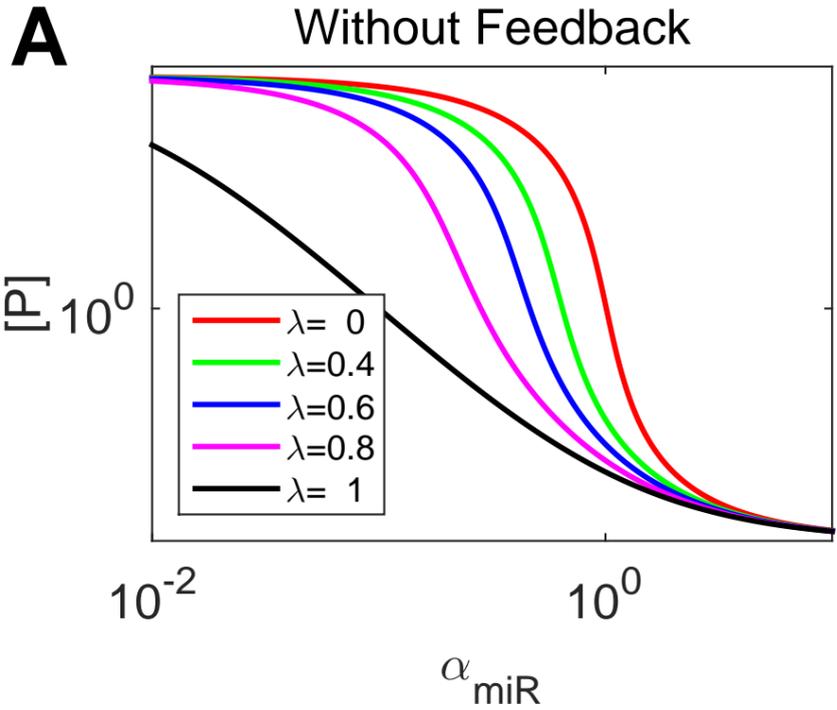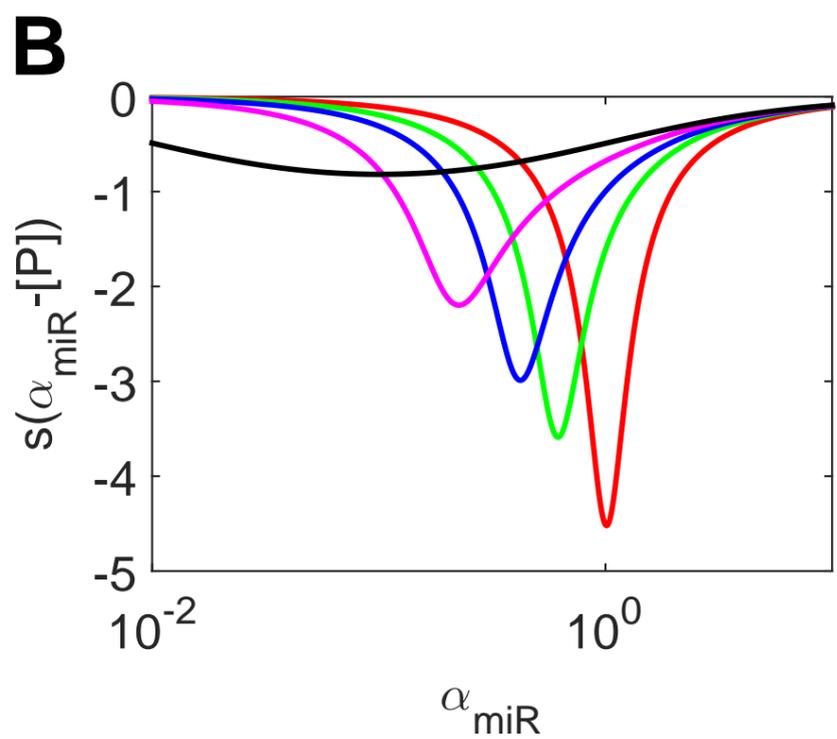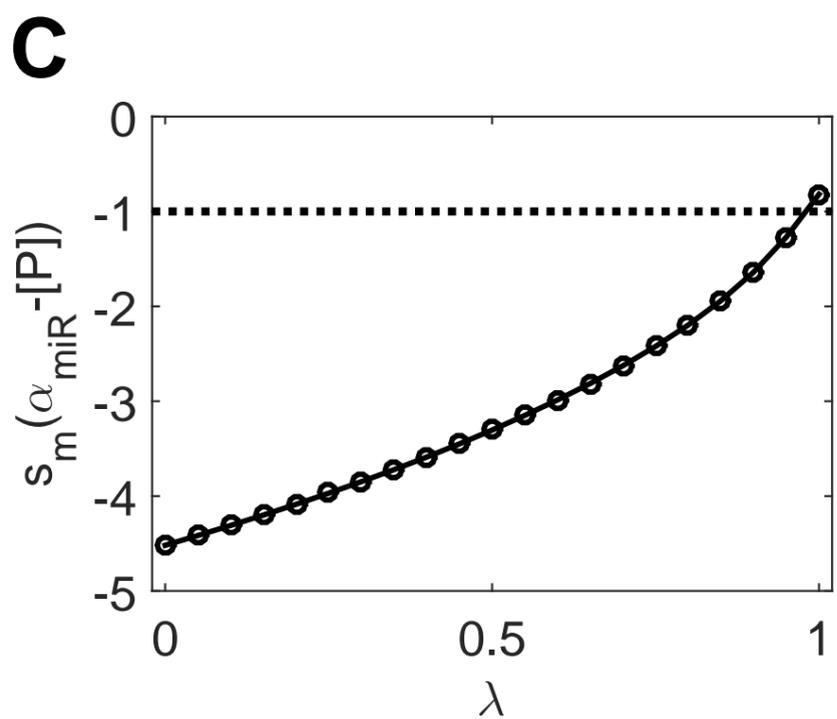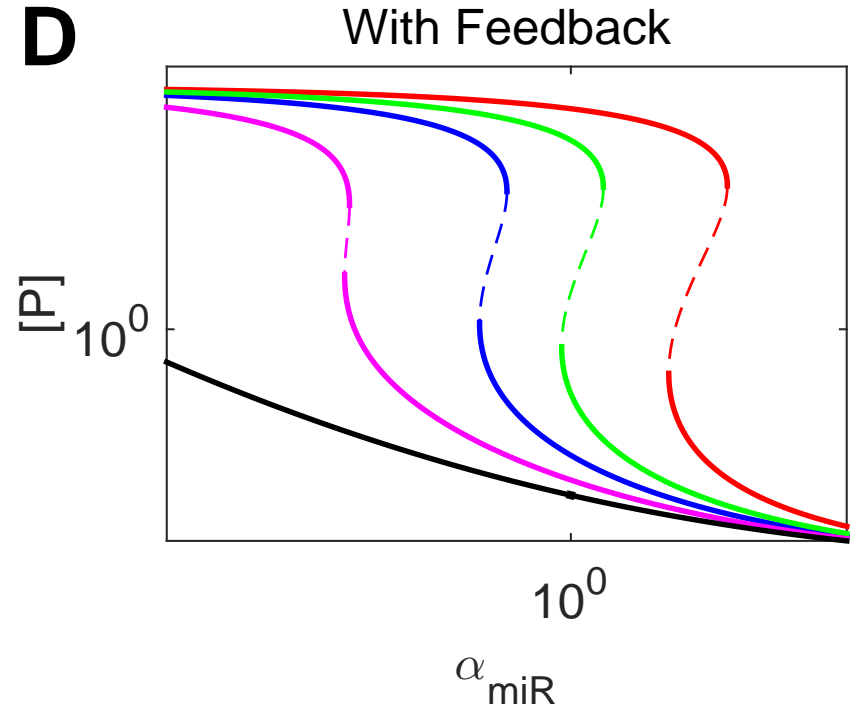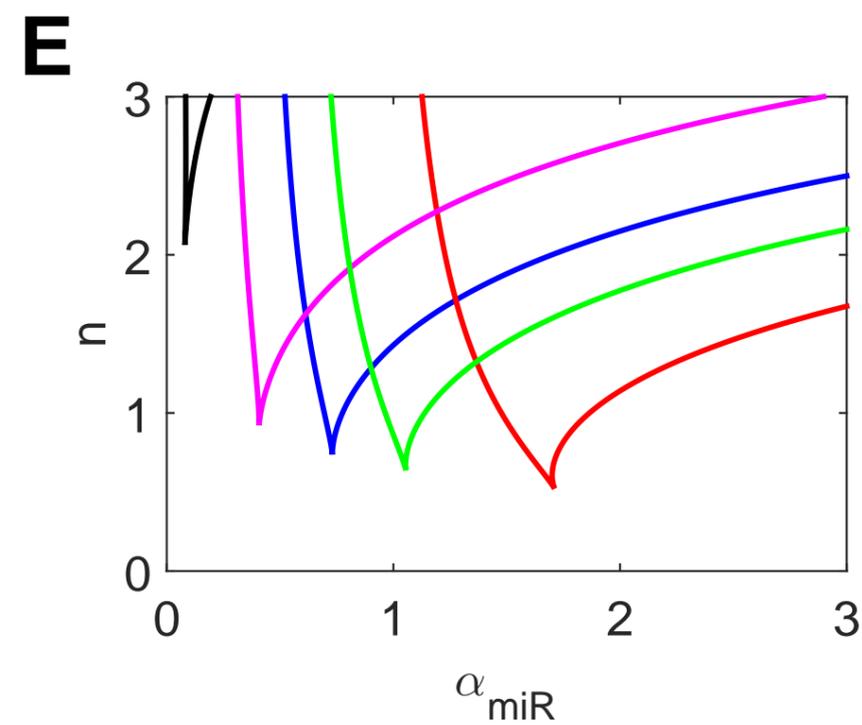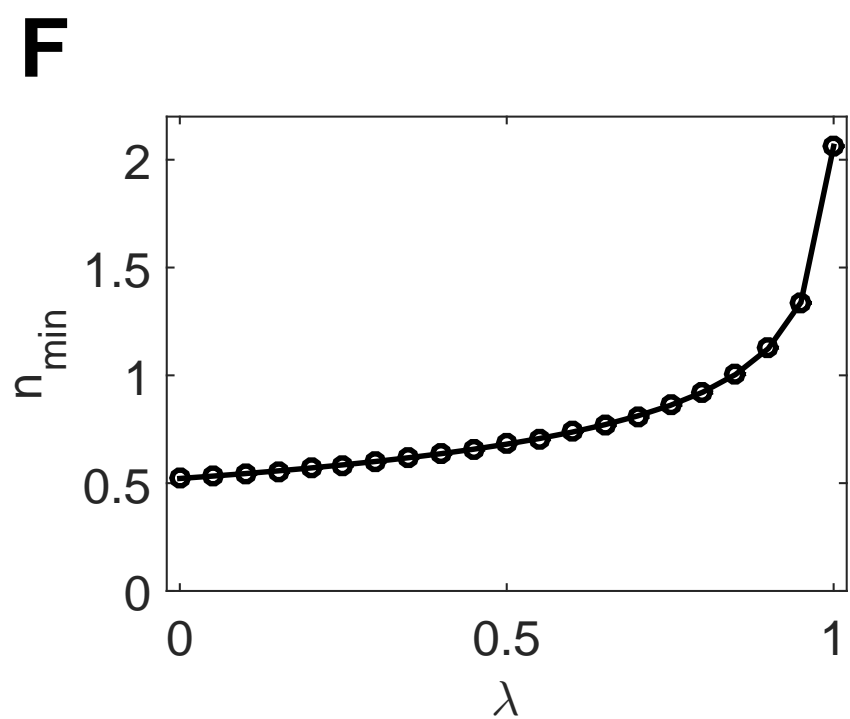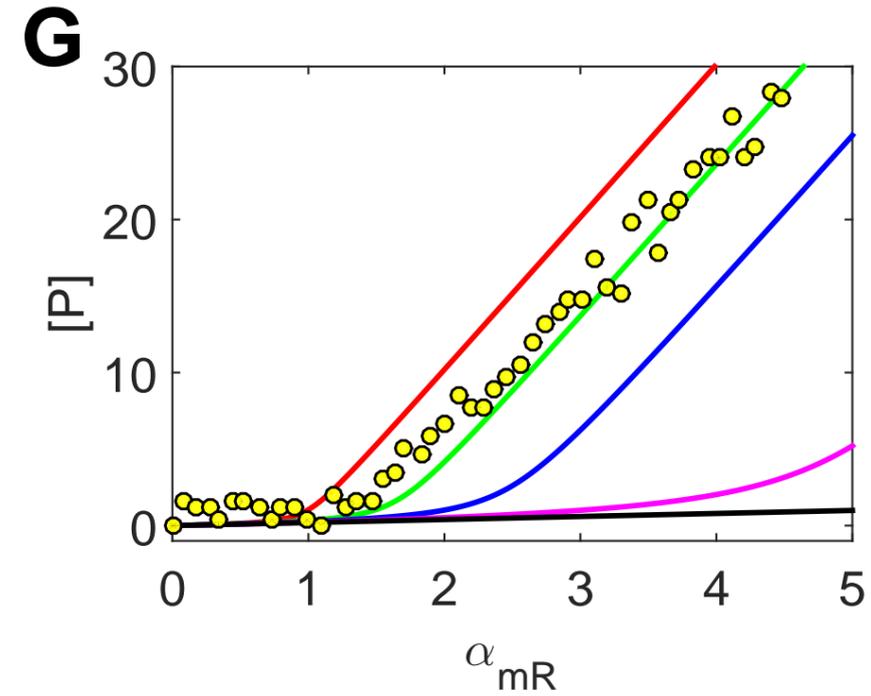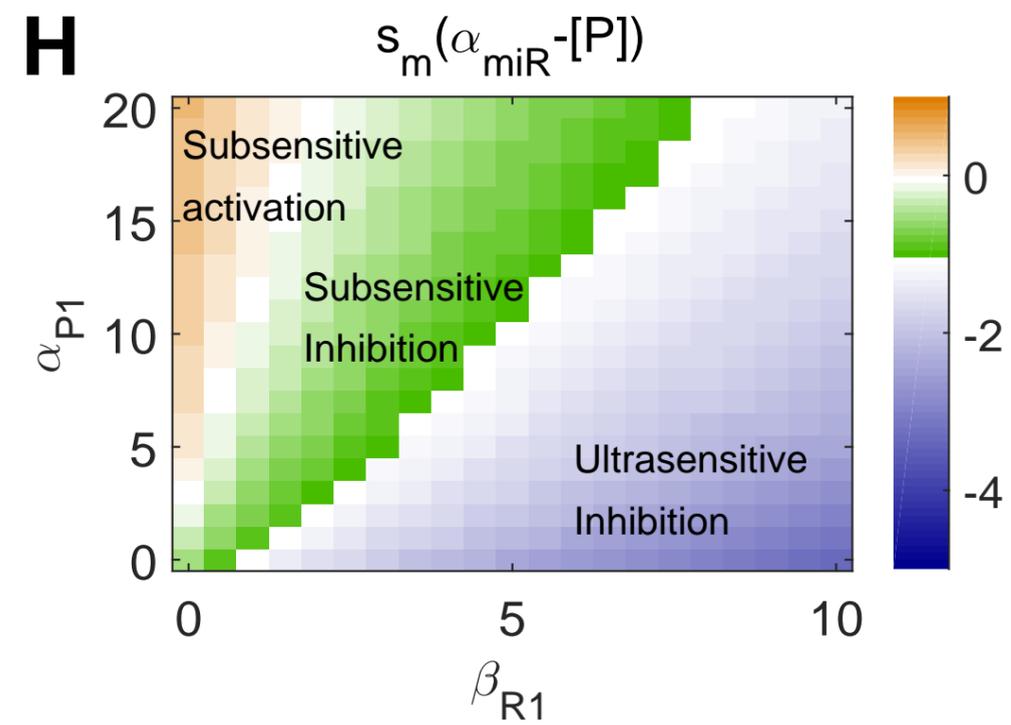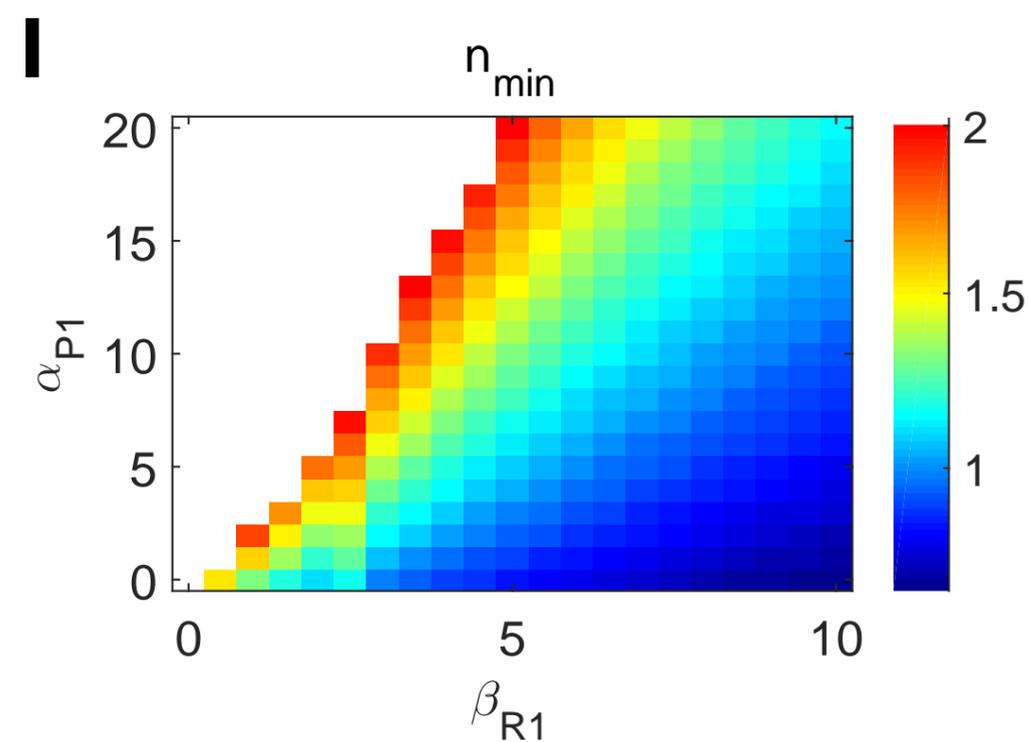

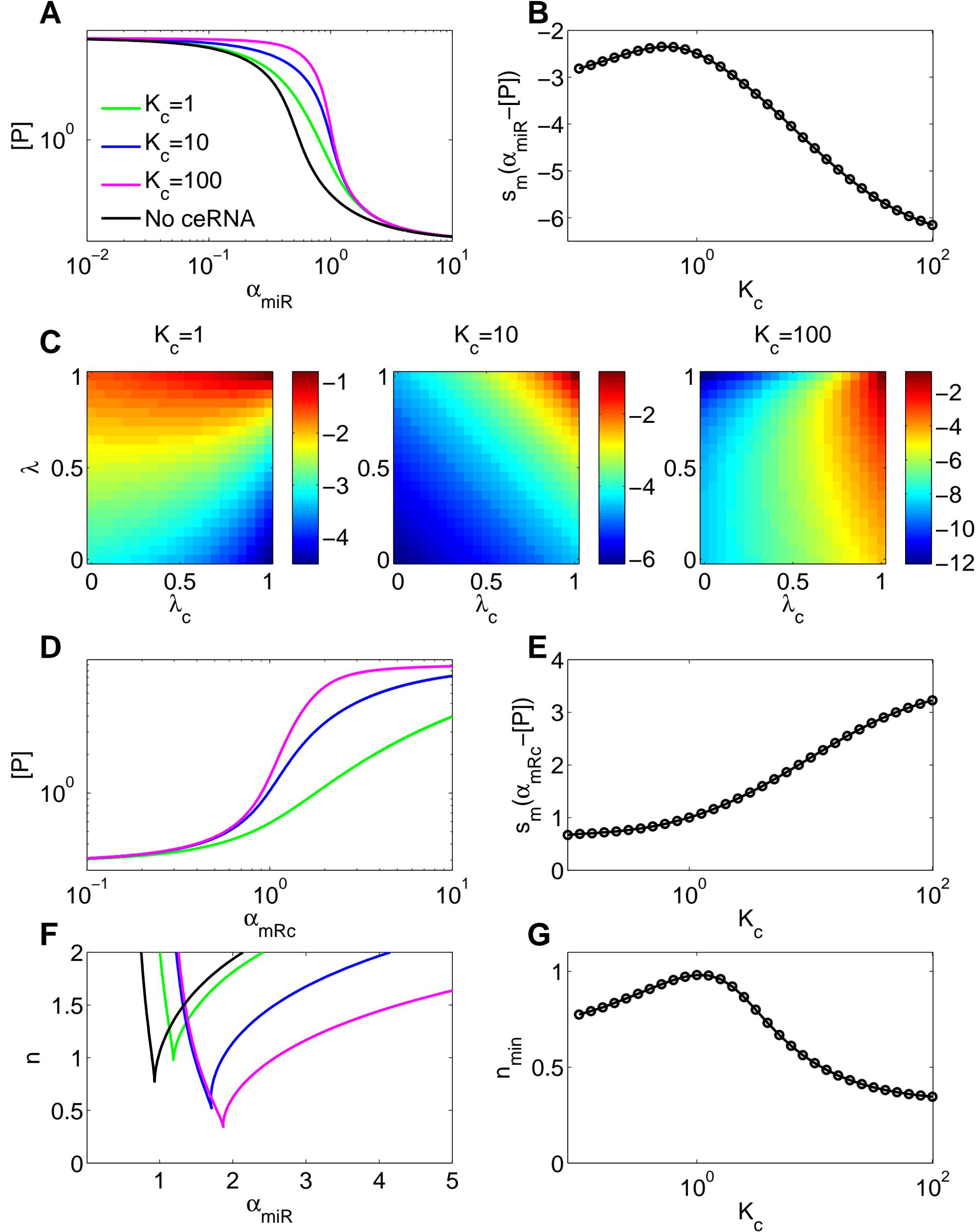

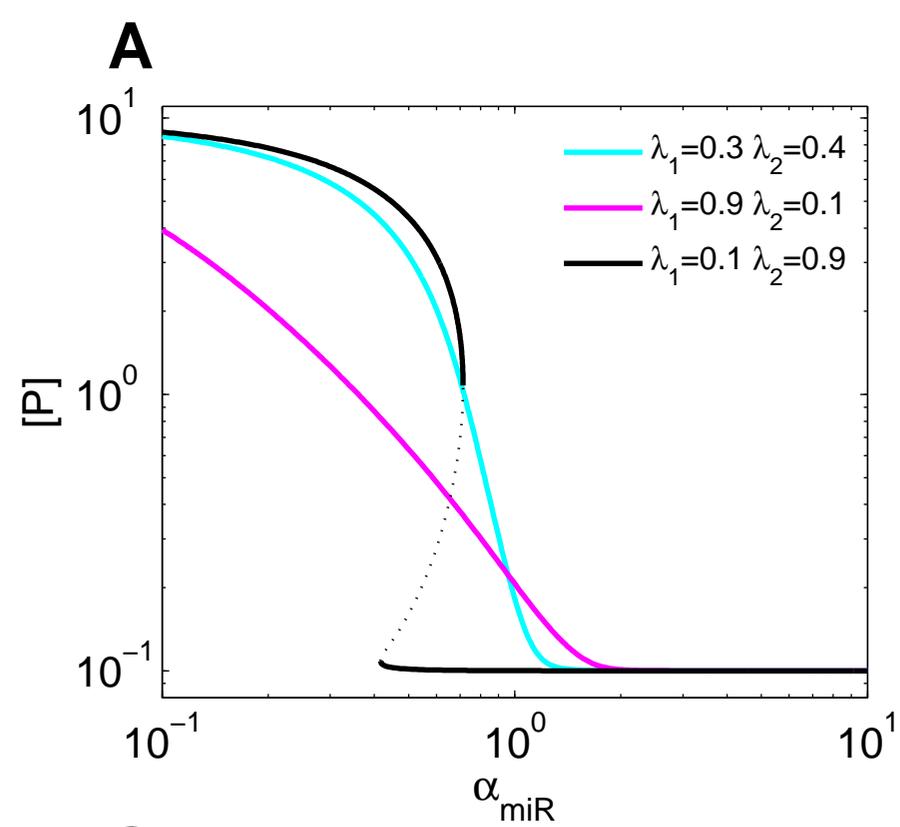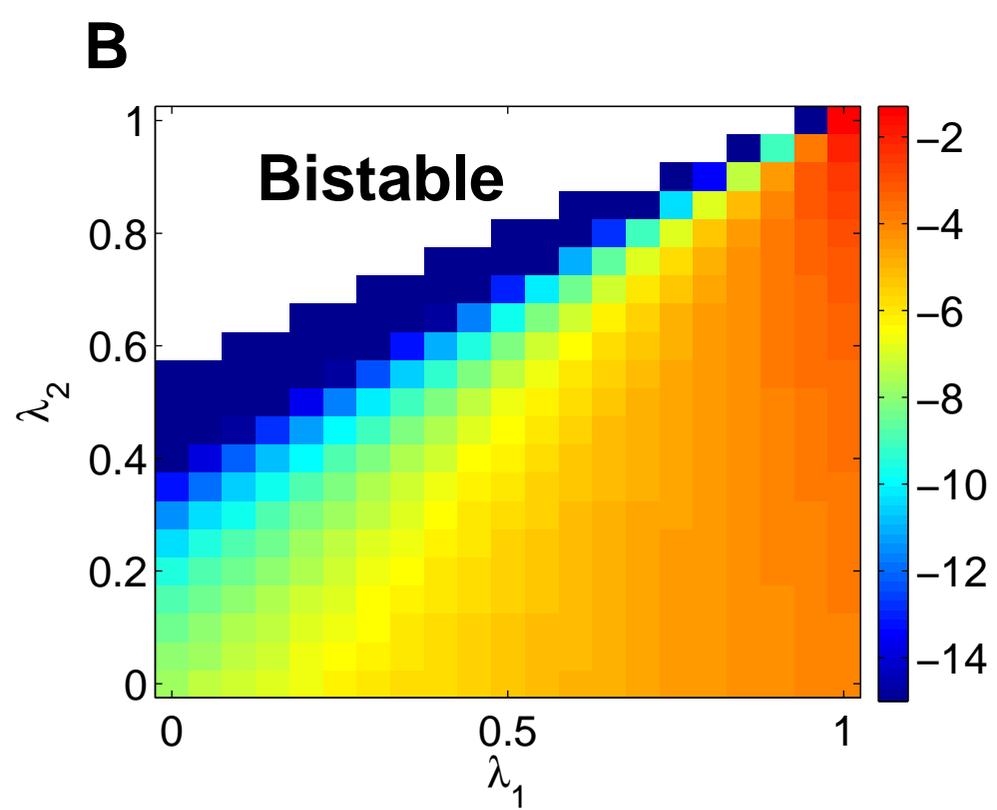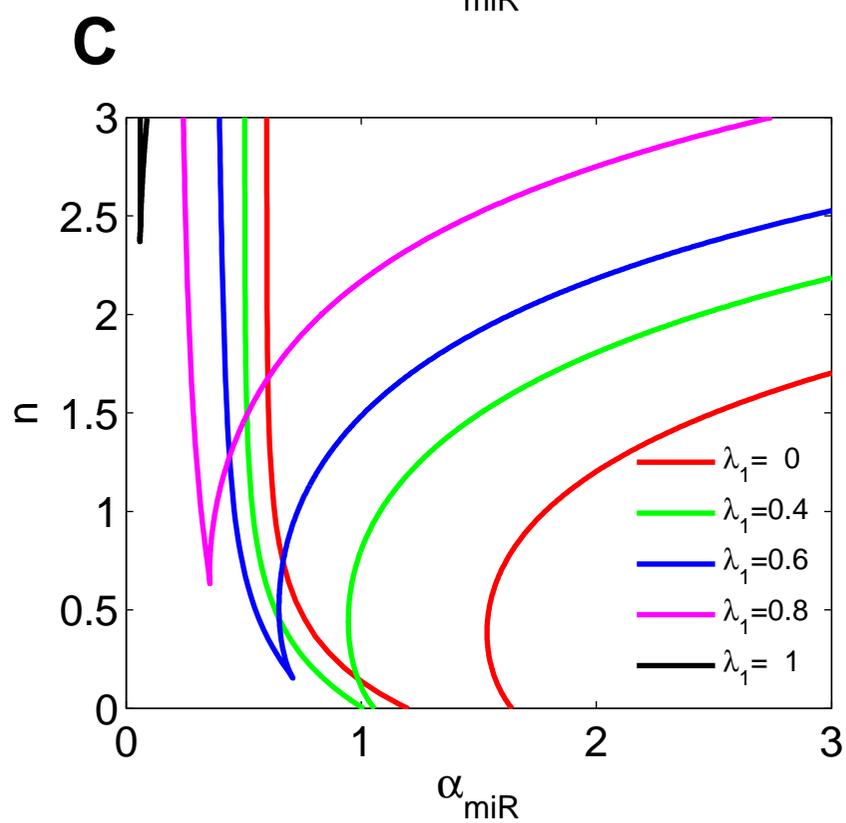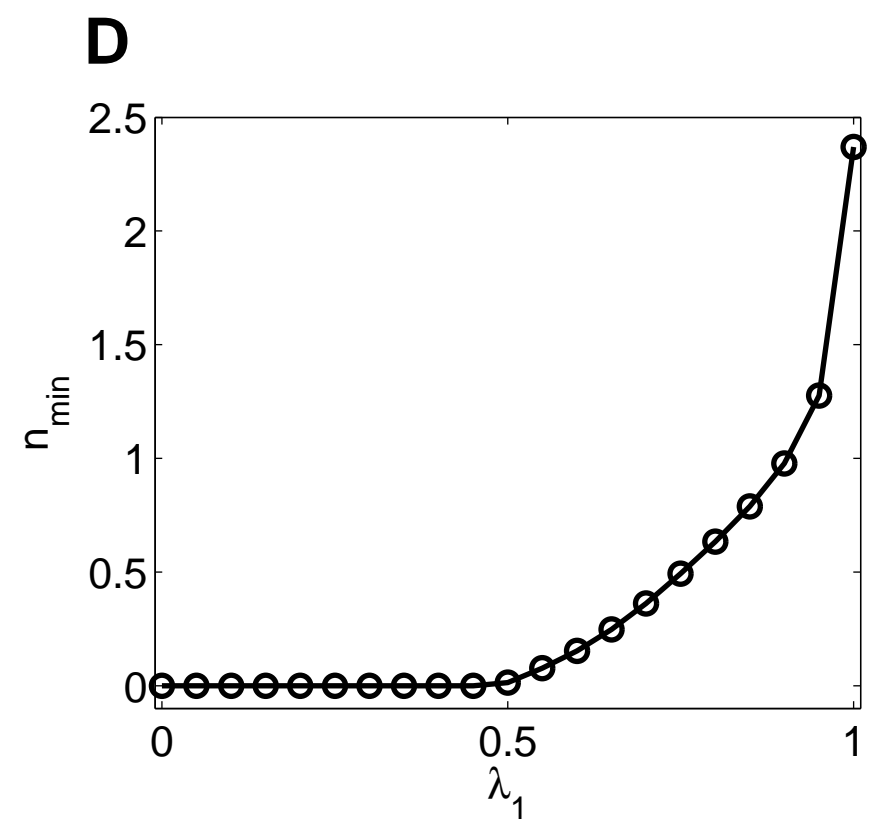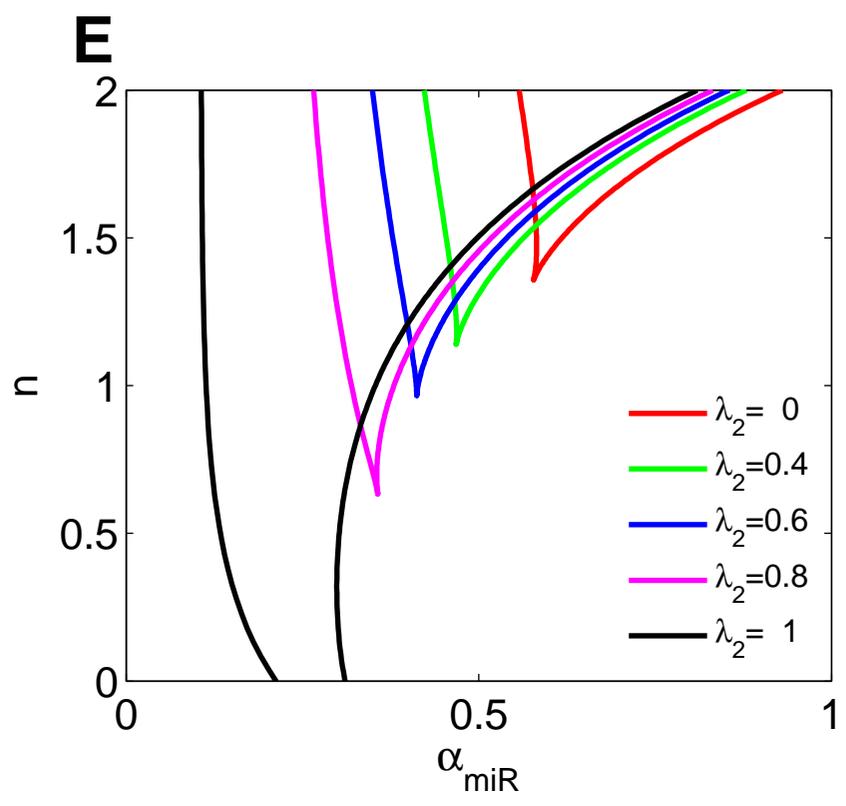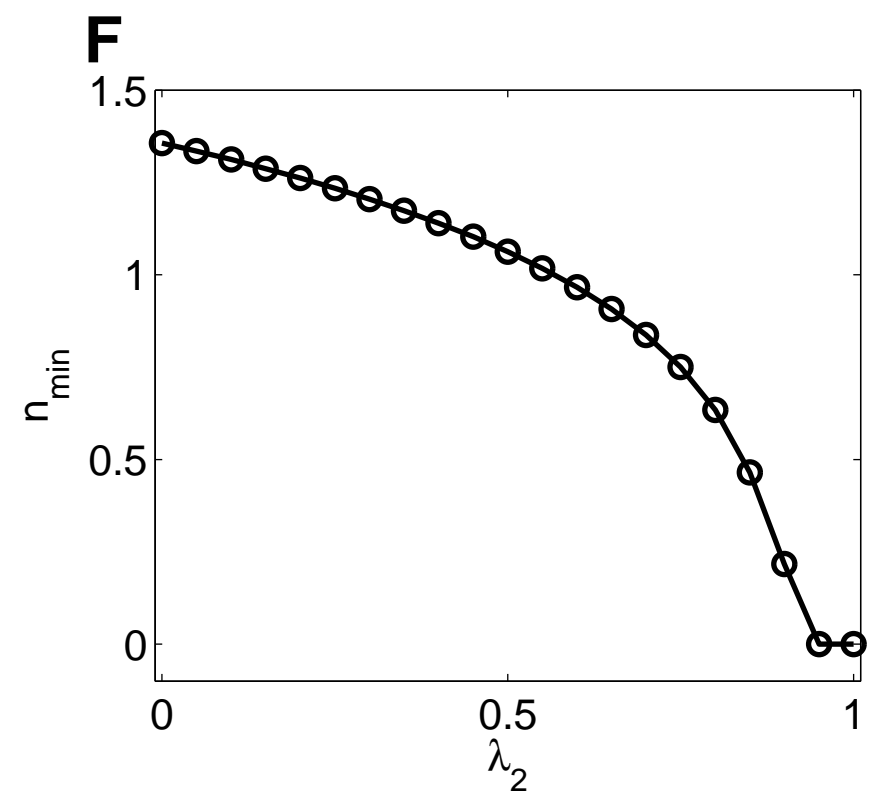

**FigS1.** Method of locating the minimum of Hill coefficient $n_{min}$ to generate bistability. (A) First, one-parameter bifurcation is analyzed to find the Saddle-Node bifurcation points. (B) Then, two-parameter bifurcation is analyzed to find the Cusp bifurcation pints, where $n_{min}$ is located.

**FigS2.** Nullclines of the total levels of protein and miRNA.

**FigS3.** Dependence of the maximum sensitivity $s_m$ and minimal Hill coefficient $n_{min}$ on other parameters, (A) K1, (B) $\beta_{R1}$, (C) $\alpha_{P1}$.

**FigS4.** Dependence of the sensitivity of ceRNA-mediated regulation of mRNA on the recycle ratios $\lambda$ (A) and $\lambda_c$ (B) under different values of the ceRNA-miRNA binding constant $K_c$.

**FigS5.** Dependence of bistability boundary in the space of $\lambda_1$ and $\lambda_2$ on the mRNA-miRNA binding constant.

**FigS6.** Effect of ceRNA on the minimum of Hill coefficient $n_{min}$ for bistability under the case with two miRNA binding sites on mRNA. (A) Two-parameter bifurcation diagrams for $\alpha_{miR}$ v.s. n respectively with different ceRNA-miRNA binding constant $K_c$. (B) The dependence of $n_{min}$ on ceRNA-miRNA binding constant $K_c$. $\lambda_1 = \lambda_2 = \lambda_c = 0.5$, $\alpha_{mRc} = 1$, $\beta_{mRc1} = 10$.

**FigS7.** Effect of ceRNA on the minimum of Hill coefficient $n_{min}$ for bistability under the case with two miRNA binding sites on ceRNA. The dependence of $n_{min}$ on ceRNA-miRNA binding constant $K_{c1}$ (A) with $K_{c2} = 100$, and $K_{c2}$ with $K_{c1} = 100$ (B). $\lambda_1 = \lambda_{c1} = \lambda_{c2} = 0.5$, $\alpha_{mRc} = 1$, $\beta_{mRc1} = \beta_{mRc2} = 10$.

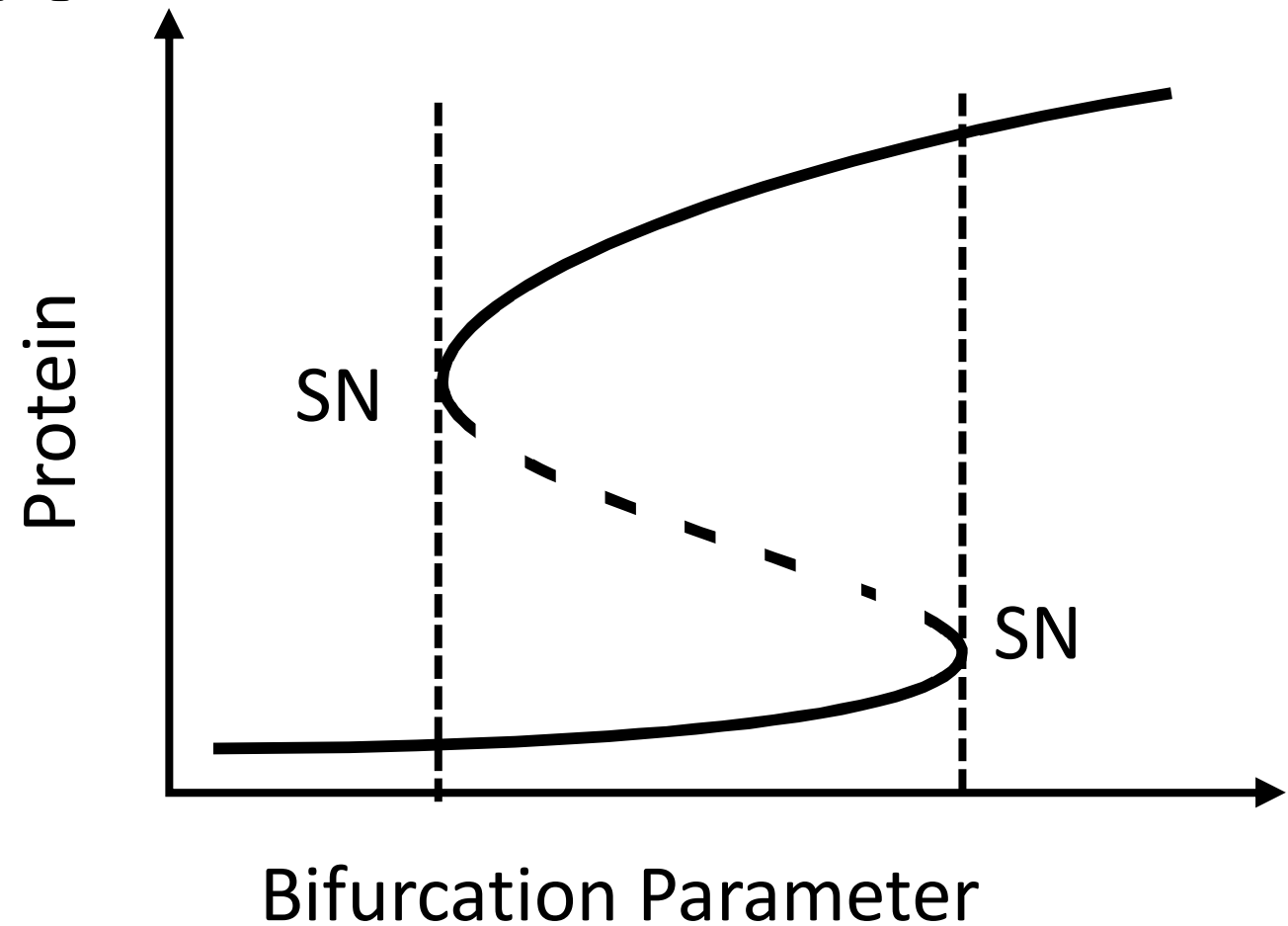 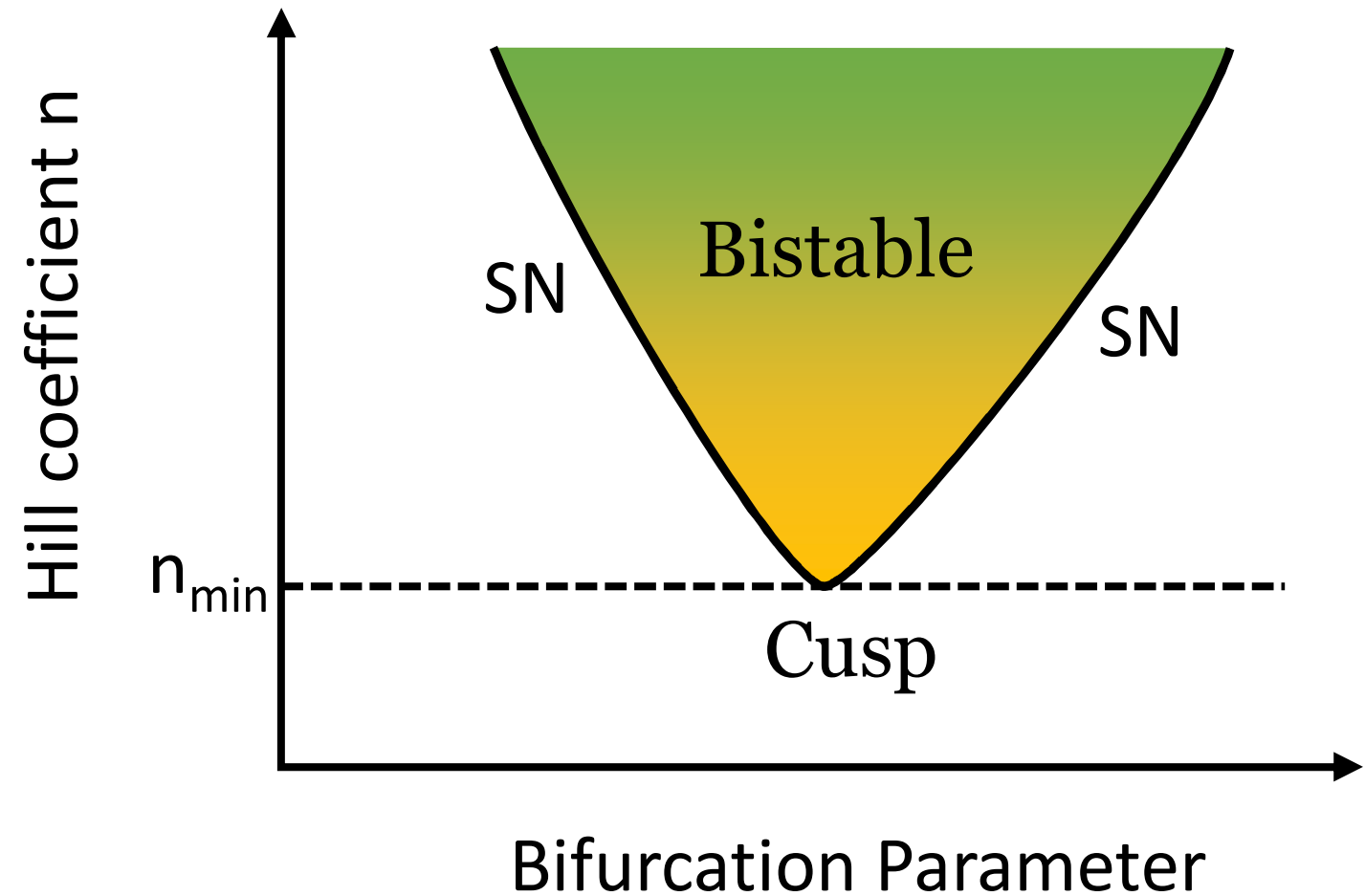

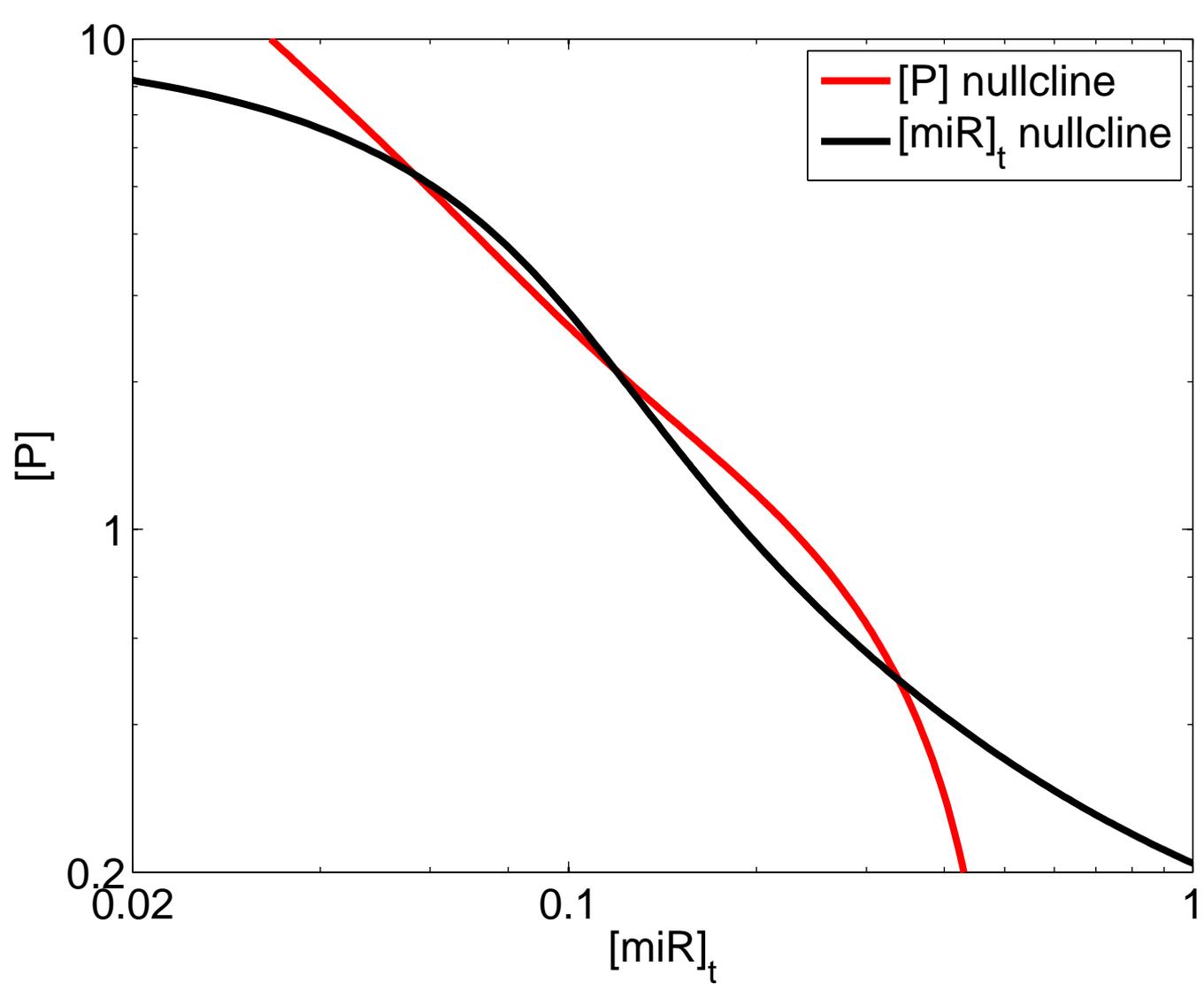

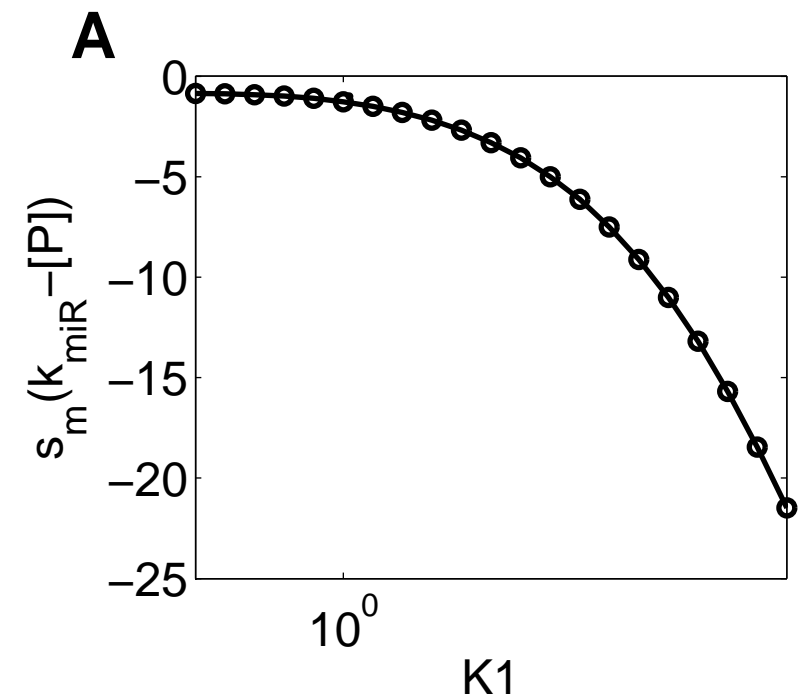 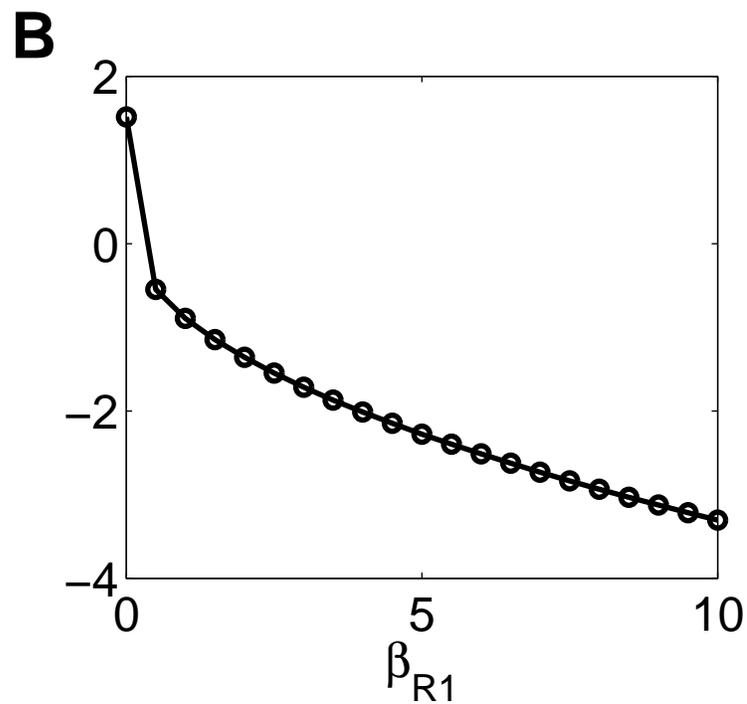 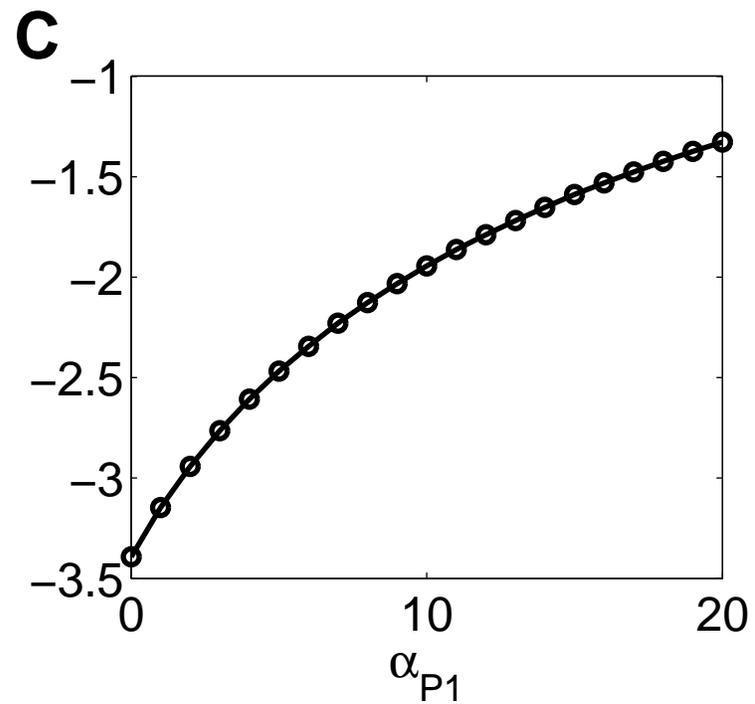
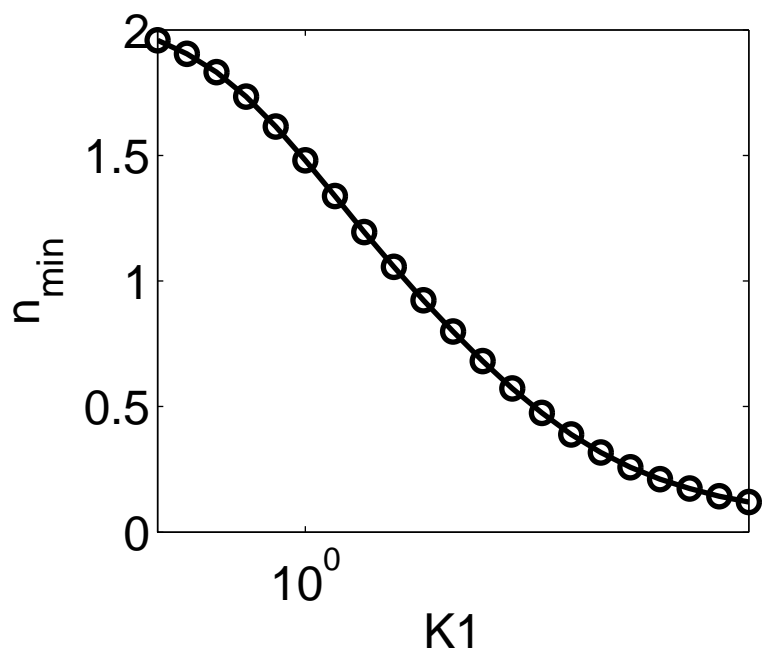 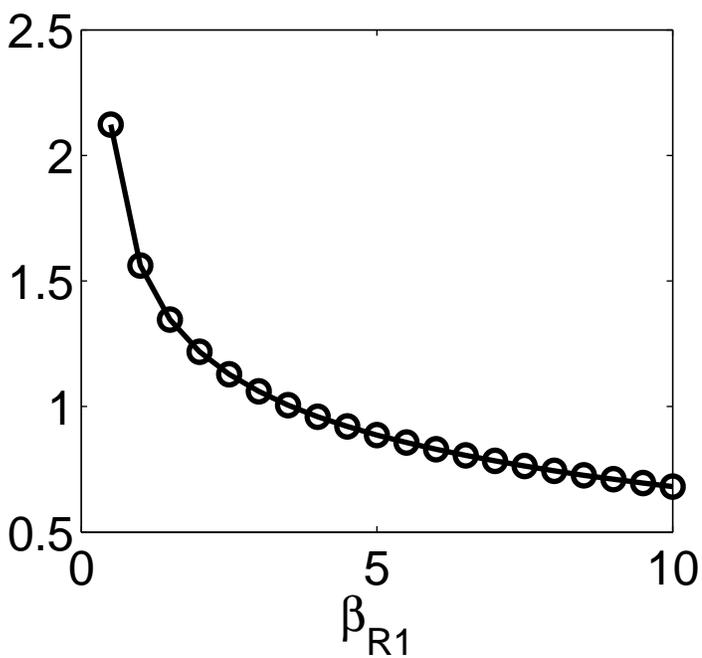 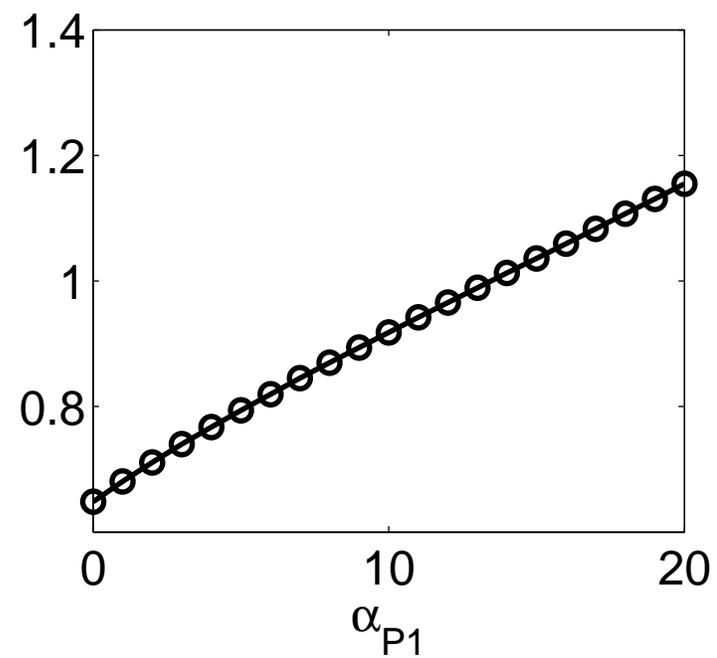

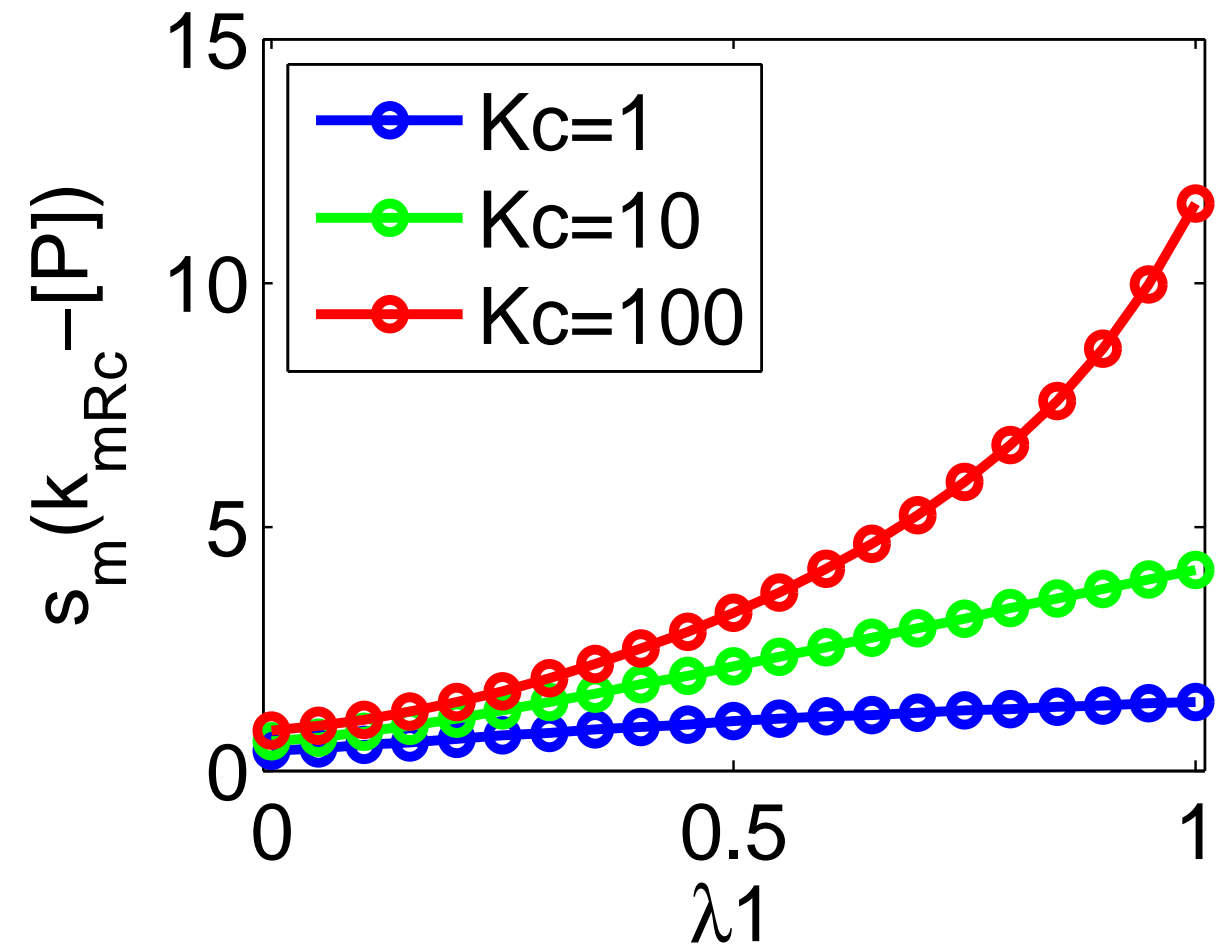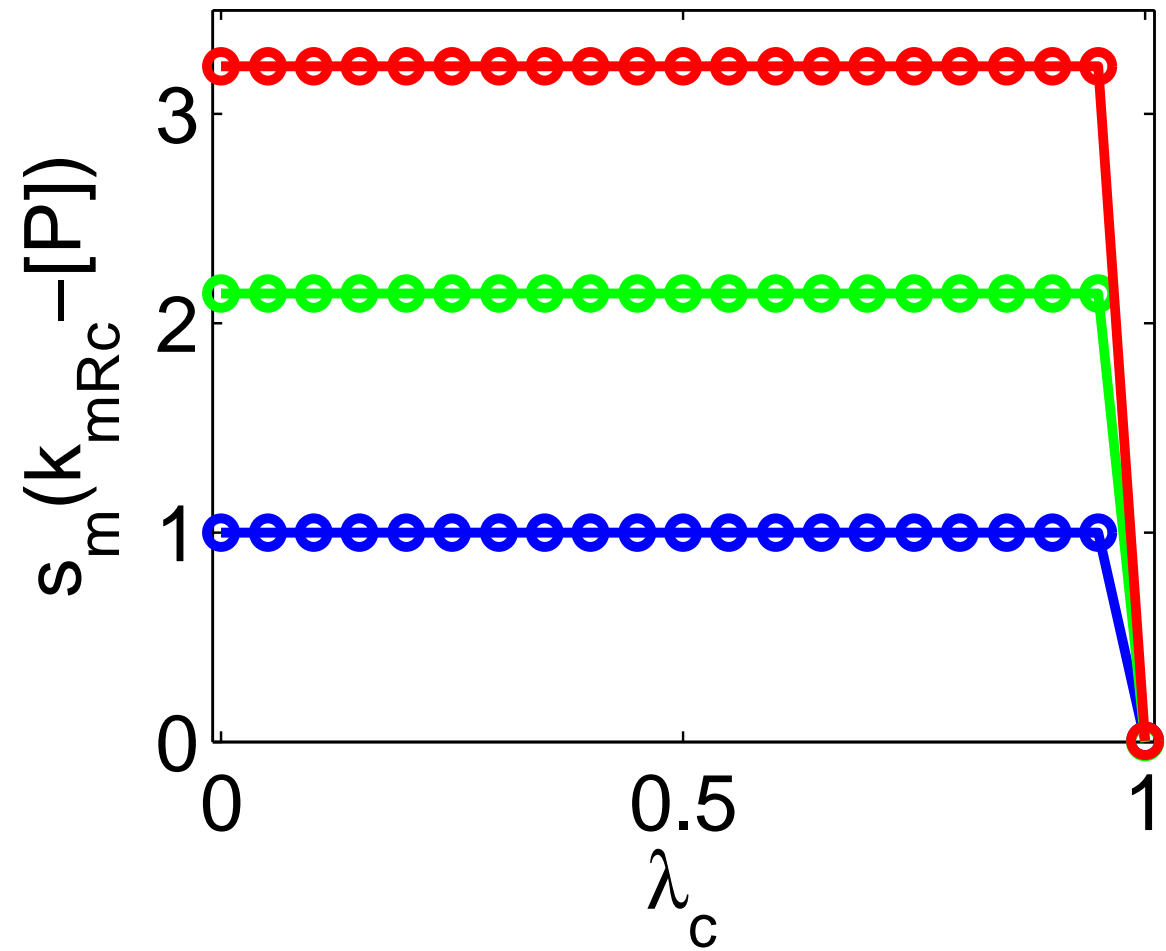

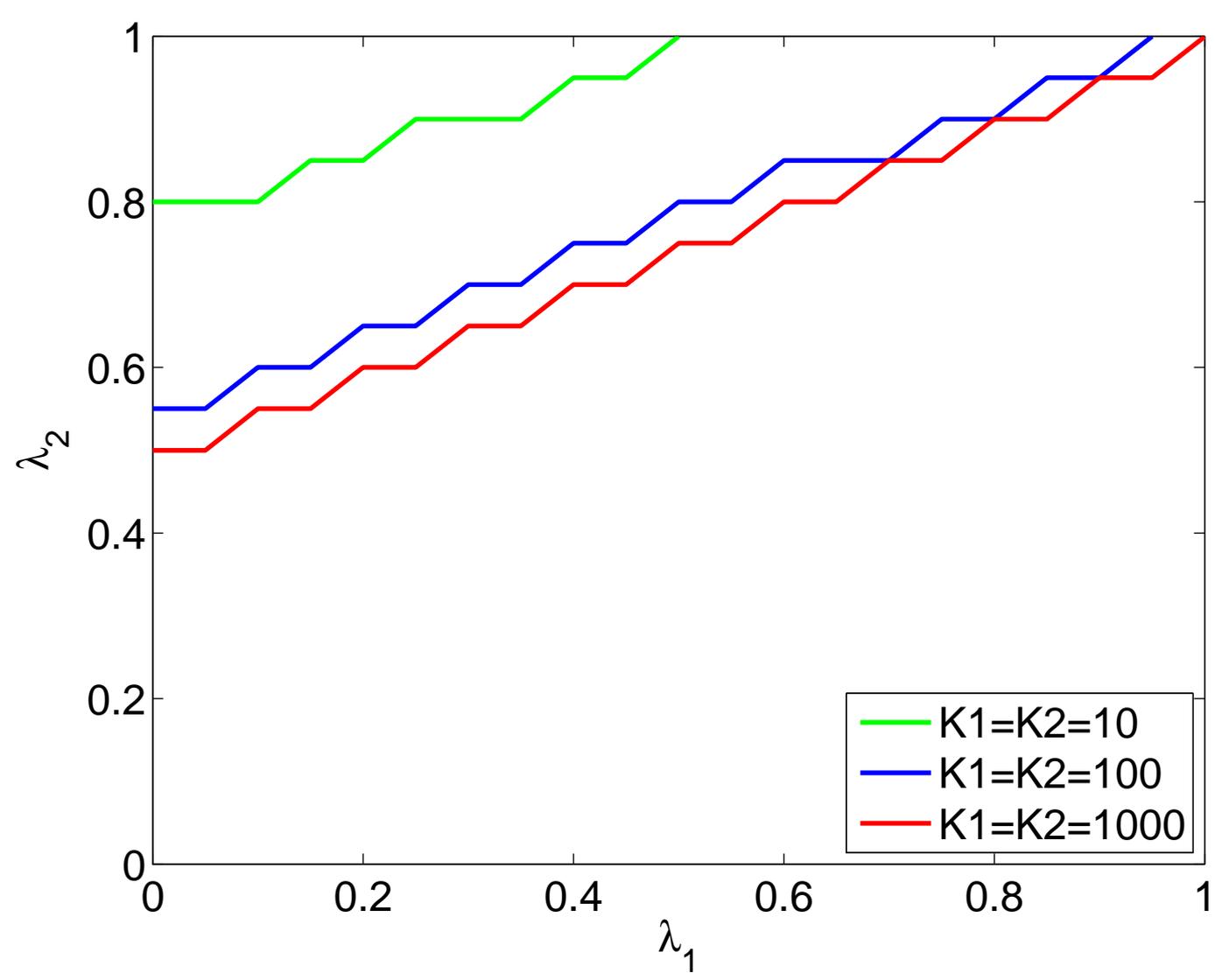

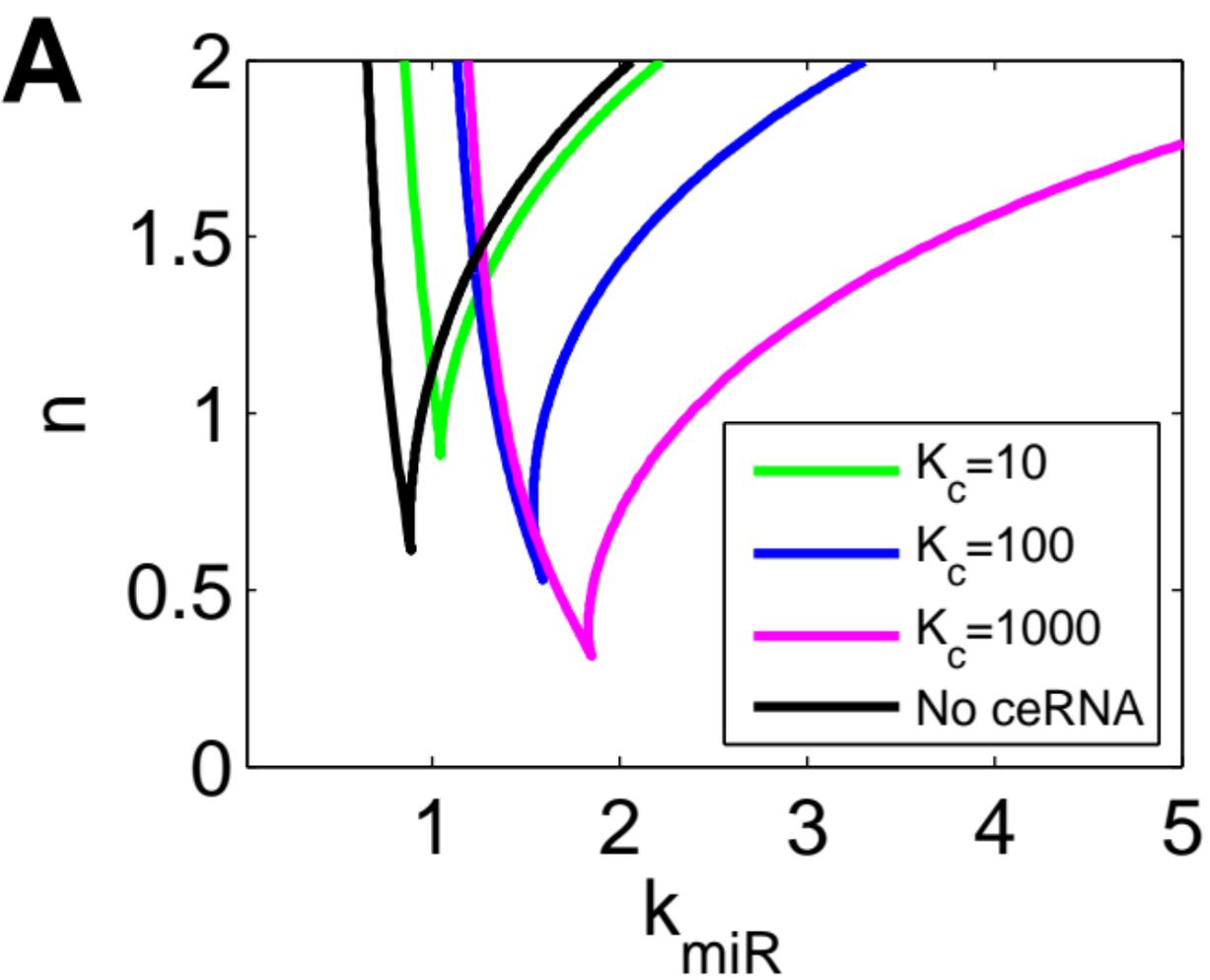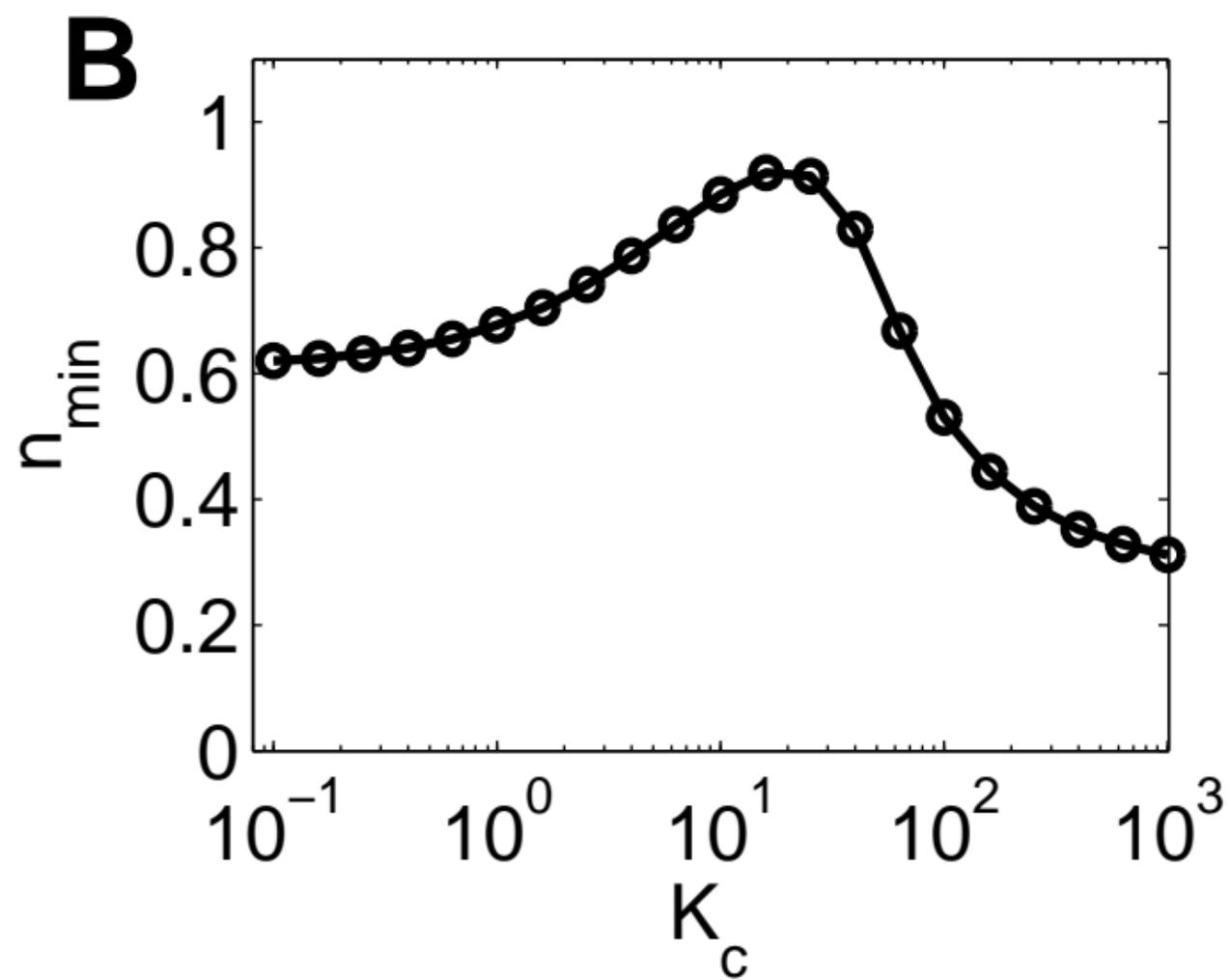

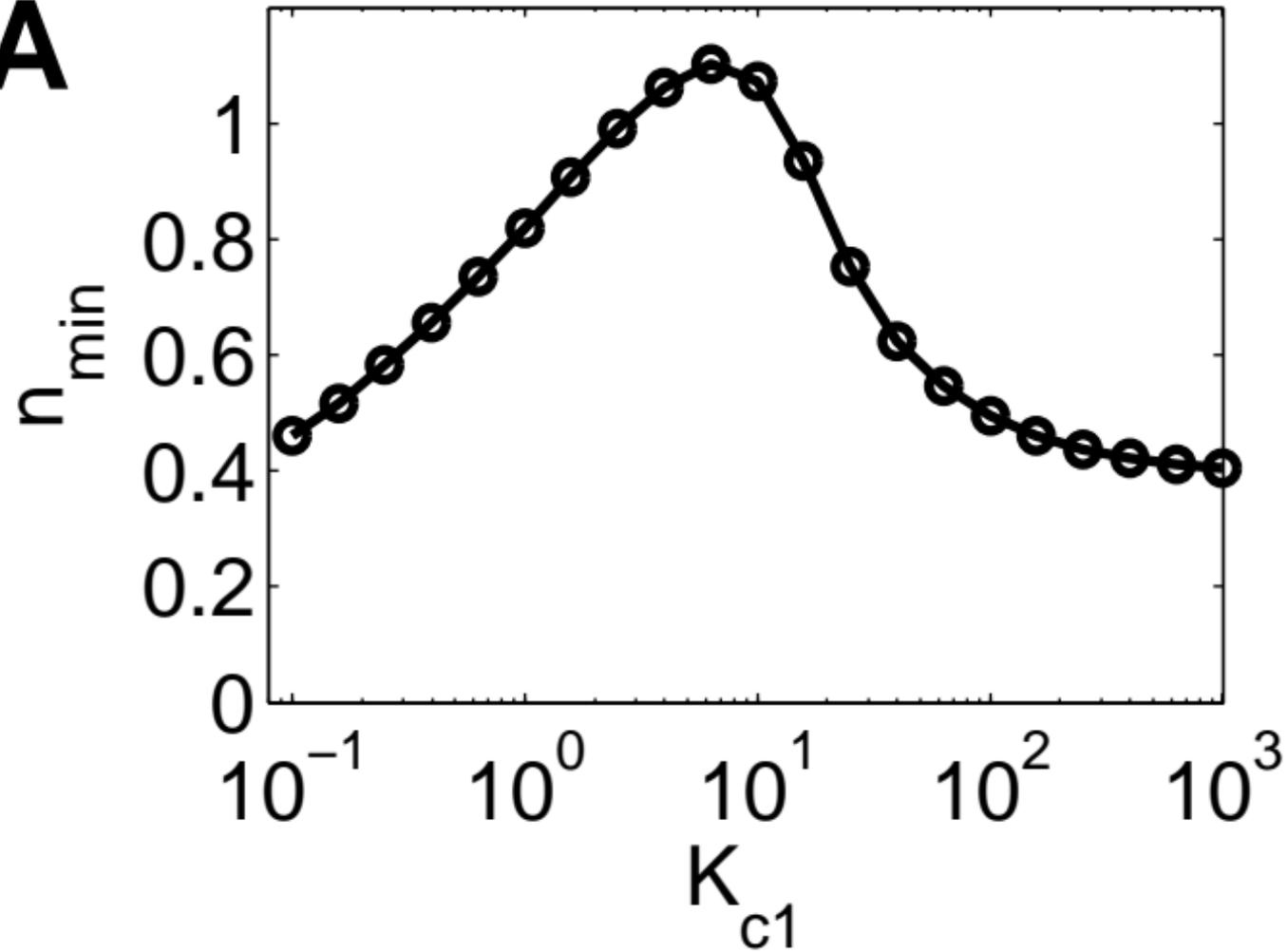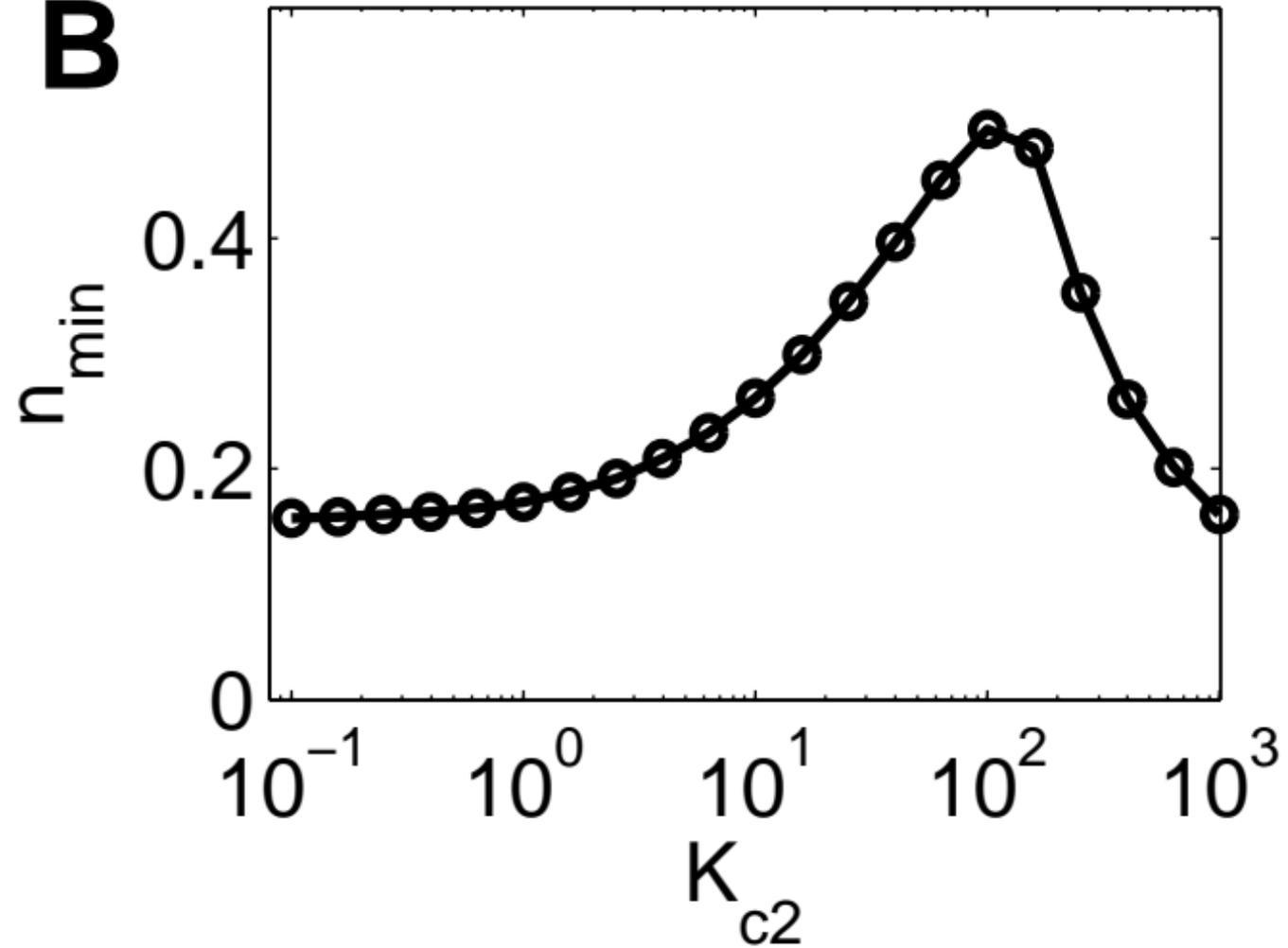